\documentclass{elsart}
\usepackage{epsf,epsfig,amssymb}
\newcommand{\R}{\mathbb R}
\begin{document}
\begin{frontmatter}
\title{
Fourier transform method for  imaging 
atmospheric Cherenkov telescopes
}
\author[UdM]{A. Atoyan,}
\author[UdM]{J. Patera}
\address[UdM]{CRM, Universit\'e de Montr\'eal, C.P. 6128 Centre-Ville,  
Montr\'eal, Canada H3C 3J7}
\author[YerPhI]{V. Sahakian}
\author[YerPhI]{and A. Akhperjanian}
\address[YerPhI]{Yerevan Physics Institute, Alikhanian Brothers, 2, Yerevan 
       375036, Armenia}

\begin{abstract}
We propose Fourier transform (hereafter FT) method for processing 
images of extensive air showers (EAS) detected by imaging 
atmospheric Cherenkov telescopes 
(IACT) used in the very high energy (VHE) gamma-ray astronomy. The method 
is based on the discrete Fourier transforms (DFT) on 
compact Lie groups, and the use of {\it continuous extension} of the 
inverse discrete transform for 
approximation of discrete EAS images by continuous EAS brightness 
distribution functions. Here we 
present the FT-method for SU(3) group. It allows practical 
realization of the DFT method for
functions sampled on hexagonal symmetry grids implemented in most 
of the current IACT cameras. We note that the proposed FT-method can also 
 be implemented for the rectangular grids   
using the DFT on $\rm SU(2)\times SU(2)$ group. 

We present the results of application of the FT-method
to Monte-Carlo simulated bank of TeV proton and gamma-ray EAS
images for a stand-alone telescope. Comparing between the
FT-method and the currently used standard method for signal 
enhancement, on the basis of parameter ALPHA, shows that
the FT technique allows a better and systematic increase 
 of the gamma-ray signal. The relative 
difference between these two methods becomes more profound especially for 
`photon poor' images, for which the standard method deteriorates. 
It suggests that the EAS detection thresholds could be effectively 
reduced with implementation of the FT technique for IACTs. 
This prediction is further supported by a significant noise suppression
capability of the method using simple `low-pass' filters 
in the image frequency domain. This new approach allows very deep `tail' 
(and `height') image cuts, differentiation of images, frequency spectra, etc., 
which can be used for
development of new effective parameters for the EAS image processing.    
\end{abstract}

\end{frontmatter}

\section{Introduction}

Since last 15 years, starting from the first confident detection 
of the Crab Nebula ~\cite{crab89},  the IACTs have proved 
powerful detectors of very high energy (VHE, conventionally 
$E\geq 100 \,\rm GeV$) gamma-rays sources. Due to IACTs, and 
in particular, stereoscopic IACT systems (see ~\cite{AhAk97}), 
VHE gamma-ray astronomy is now an established 
and rapidly growing observational astronomy, with proven  
 experimental methodology and with already significant number of 
detected sources of various types (see ~\cite{TW03} for the 
comprehensive recent review). As well known,
the power of IACTs is based on the   
realization of the idea ~\cite{tw78} that extensive air showers (EAS) 
induced by cosmic rays (CRs) and gamma-rays can be 
effectively discriminated from each other exploiting the intrinsic differences 
between their Cherenkov light images detected by the multichannel optical 
photoreceivers (cameras) of the telescopes.
 
The standard method (hereafter, S-method) for such image 
discrimination involves various 
"shape" and "orientation" parameters proposed first by Hillas 
~\cite{hillas}, which are based on the first and  
second-order moments of the EAS image $\{G_{J}\}$. Recall that the  
EAS image is a set of numbers $G_{J}$ of photoelectrons registered by the 
$J$-th photomultiplier tubes (PMTs) of the IACT camera. The integer index
$J=1,2,... , K$ corresponds to the discrete 2-dimensional spatial 
({\it angular}) coordinates of the PMTs, $J = (j,m)$.   
In order to improve the performance of the `standard parameterization', 
 different image preprocessing 
 techniques have been suggested (e.g. ~\cite{mohanty,lessard,bhb}). 
Note that all of these image filtering methods are based on the 
image processing directly in the coordinate (i.e. {\it image}) space.

Meanwhile, one of the known powerful methods used for signal 
processing generally is the method of discrete Fourier transform 
(DFT; see e.g. ~\cite{oppenh,lj90}). In this method the discrete image 
is transformed into wave number (or frequency, for transformation 
of the time argument) space, in which different `filters' can be applied. 
Then the filtered image 
 is recovered by the inverse transform. Yet, this approach has not been 
implemented for the EAS images detected by IACTs. Perhaps, it could be partly 
explained by the hexagonal/triangular symmetry of the PMT grids implemented 
in most of the current IACTs, 
whereas the known methods of DFT  are 
developed for images defined on grids of rectangular symmetry.   
  
Here we propose a method which allows DFT  
of 2-dimensional images sampled on hexagonal grids. This technique 
represents a particular case of implementation of a general method for  
Fourier analysis of multidimensional discrete functions 
which uses the so called orbit functions of compact Lie groups 
as the transform basis ~\cite{MP1,MP2,P03}. To distinguish it from 
standard DFT, we abbreviate the DFT on Lie groups
as DGT, standing for {\it discrete group transform}.
Images defined on grids with hexagonal symmetry 
belong to the case of Fourier transform on the group SU(3), whose root 
lattice displays that symmetry. 
In Section 2 we show that the method 
is rather simple and can be used in practice even without 
any specific knowledge of the Lie group theory. 

Note that the case of DGT on the group SU(2), which has been considered 
in detail in ~\cite{CEDCT}, is reduced to a specific type of DFT 
known earlier as {\it discrete cosine transform} (DCT, see ~\cite{DCT}). The 
2-dimensional version of DCT is notorious as the basis for the high-performance
image compression standard JEPG. It corresponds to DGT on the 
Lie group SU(2)$\times$SU(2), and allows processing of images sampled 
on the ordinary rectangular grids. Thus, although in this paper we 
describe the Fourier
transform method (hereafter,  FT-method) 
for the IACTs with hexagonal 
symmetry of the PMT grids used in most of current IACTs
~\cite{hess,MAGIC,veritas,cangaroo3}, 
the FT-method is equally applicable for IACTs with the grids of rectangular 
symmetry (e.g. ~\cite{cangaroo2}).

Our previous study of DGTs has shown that these 
transforms are distinguished by some valuable 
properties. In particular, the {\it inverse} DGTs can be treated as
continuous Fourier polynomials, or {\it continuous extensions} (CE) 
of the discrete transform which interpolate well the discrete 
image at any point between the grid. Unlike 
continuous extensions of the standard DFT, but very similar to the canonical
{\it continuous Fourier transform} polynomials, the CEDGTs are 
converging functions. Furthermore, the known properties of {\it localization} and 
{\it differentiability} of the continuous Fourier transform series hold 
for CEDGT polynomials as well (see ~\cite{CEDCT}). 
 
After describing in Section 2 the basic mathematical technique of the 
Fourier transform on the Lie group SU(3) and demonstrating the ability
of CEDGT to approximate the original 
analytic functions, in Section 3 we 
apply this technique for interpolation of Monte-Carlo (M-C) simulated 
images of gamma-ray and proton induced EAS. The goal of this
paper is first of all to present the Fourier transform method itself
and to discuss some of the related new possibilities for the IACT image 
processing, rather than achieving the best result for the contemporary 
stereoscopic IACT arrays.  
Therefore here we use EAS images simulated for a stand-alone IACT 
with parameters for both the camera and the collector
corresponding to those of the HEGRA telescopes (see ~\cite{HEGRA}). 
It assumes a hexagonal 271-PMT camera with 
angular size of PMTs $h = 0.25^\circ$, and the telescope   
collector area $\simeq 8.5 \,\rm m^2$. 
These relatively modest parameters result in  comparatively coarse (`poor') 
images. For such images the performance of the proposed FT-method, 
which consists in Fourier transformation
and then interpolation of  discrete images by CEDGT,
is significantly better than the performance of the S-method.  

In Section 4 we discuss some possibilities for image processing 
in the frequency domain that can be opened with the use of 
the FT-method. These possibilities include `denoising' of the image 
by cutting off the highest frequency harmonics before the use of CEDGT. 
This is a very simple low-pass filtering procedure which, nevertheless, 
appears
rather effective for practical purposes. We will also demonstrate here,
although only qualitatively rather than quantitatively 
(which would require a separate detailed study) that the proposed 
``functional'' approach contains a significant potential for reducing 
the effective energy threshold of the IACTs.

Throughout this paper the comparing with the S-method 
is done  only on the basis of parameter ALPHA which 
has proved to be a very efficient single parameter 
used in the `standard parameterization'.
In order to keep this comparison straightforward, in this paper we 
refrain from introducing new parameters for image discrimination, 
which in principle would be possible having a continuous function 
instead of a discrete image.

 \section{Fourier transforms on SU(3) }
 
\subsection{SU(3) Fourier transform on a grid}

In general terms, the concept `Fourier transform on a grid' means that a 
discrete function $\{ G_k \mid k=0, 1,..., K \}$ defined on the vector
points $\{ {\bf r}_k \mid k=0, 1,..., K \}$  of a grid in $\R^2$ 
(or generally in any $n$-dimensional real space $\R^n$), can be 
represented as a linear combination of 
wave functions
$e^{i ({\bf k}_i \cdot {\bf r})}$ sampled on the same grid.
 The set of wave numbers 
(or `frequencies' in case of time variable) involved is finite, i.e. 
$\{ {\bf k}_i \mid i = 0, 1, ... K^\prime\}$, where $K^\prime$ is not 
necessarily equal to $K$. It is important, however, that
the action of the transform operator $\hat T$
would produce an unambiguous `image' of $\{ G_k\}$ 
in the wave vector space,  $\hat{T} \, : \, \{G_k \}  \rightarrow 
\{ A_j \mid j= 0, 1, ... , K_1\}$ where $K_1=K$. 
Since the transform operator is linear, the numbers of
independent elements in the original and transformed images should be 
equal. 
    
The method of DFT on compact 
Lie groups, or the DGT, is based on two principal
ingredients suggested in  ~\cite{MP1,MP2} which allow practical
computation of Fourier transforms on Lie groups. These 
are the discretization of the space of variables in accordance with 
the {\it elements of finite order} (EFO) of the group, 
and the use of {\it orbit functions} 
 of the group as the transform basis.    
The basic property that makes the method work, is  
the orthogonality of the orbit functions on the sets of EFOs 
(see ~\cite{MP1,MP2,P03} for details). The EFOs of adjoint 
orders dividing an integer $N$ make a discrete grid of a definite 
symmetry (depending on the group) 
in the so called fundamental region of the group.

Without going into specific mathematical details, which are being presented
elsewhere, but in order to show the connection between the transform basis
and the Lie groups, we remind that the group SU(3) can be faithfully 
represented as the group of all 
unitary matrices of size $3\times 3$. 
Any unitary matrix can be  diagonalized by a unitary
transformation. Therefore every element of SU(3) is conjugate to at
least one diagonal element $\Theta$ in this 3-dimensional matrix
representation. The set of all such diagonal matrices forms
the {\it maximal torus} {\sf T} of the group:
\begin{equation}\label{torus}
{\sf T} = \left\{ \Theta(x,y,z) = \left( 
        \begin{array}{ccc}
          e^{2\pi i x} & 0 & 0 \\
         0  & e^{2\pi i y} & 0 \\
         0 & 0 & e^{2\pi i z}
         \end{array} \right) \; \mid \; x+ y +z = 0  
    \right\} \; .
\end{equation}
The condition $x+y+z =0$ describes 
a plane on which the unitary condition $\det {\Theta} =1$ 
for the matrices is satisfied. 
Every element of the group can be thus represented by a 
2-dimensional vector point on this plane. In particular, the {\it 
simple roots} 
of SU(3), given by vectors $\alpha_1 = (1,-1,0)$ and 
$\alpha_2 = (0,1,-1)$, correspond to the 2-dimensional vectors with lengths
equal to $\sqrt{2}$ at an angle
$120^\circ$ to each other. In that plane, the maximal torus {\sf T} 
of the group corresponds to the 
hexagonal region shown in Fig.~1a.

Two elements in {\sf T} are conjugate, if and only if they differ by 
permutation of their diagonal entries. The conjugacy 
classes in SU(3) are in 1-1 correspondence with the 
points in the {\it fundamental domain} {\sf F} of the group. 
In Fig.~1 it corresponds to the triangular segment enclosed 
between vectors $\omega_1$ and $\omega_2$ representing the
{\it fundamental weights} of SU(3). These vectors are expressed through 
$\alpha_1$ and $\alpha_2$ as 
$ \omega_1 = (2\alpha_1 +\alpha_2)/3  $ and 
$\omega_2 = (\alpha_1 + 2 \alpha_2)/3$.  
The weights $\omega_1$ and $\omega_2$ are orthogonal to 
$\alpha_2$ and $\alpha_1$, respectively,
with the scalar product $(\omega_i \, , \alpha_j ) = \delta_{i,j} $, 
 where $\delta_{ij}$ is the Kroneker symbol.

 Any vector $\lambda$ in the plane can be expressed through fundamental weights
 as $\lambda = a \omega_1 + b \omega_2 $. 
Action of the Weyl group $W$ on $\lambda$, which in case of SU(3) group is
reduced to reflections of $\lambda$ in the lines of vectors $\omega_1$ and
$\omega_2$, produces a finite set $\{W(\lambda)\}$ of vectors 
equidistant from the origin. This set represents the {\it Weyl group orbit}
 of $\lambda$ containing the following 6 elements $\mu$:
\begin{eqnarray} \label{Wlambda}
W(\lambda) & = & \{ a \omega_1 + b\omega_2 \: , \: 
 b\omega_1 - (a+b) \omega_2 \: , \:   -(a+b)\omega_1 +a \omega_2  \: ,
 \nonumber \\
  & & \; \: -a \omega_1 +(a+b) \omega_2 \: , \: 
(a+b)\omega_1 - b \omega_2 \: , \: -b\omega_1 - a \omega_2 \, \}
\end{eqnarray}
In particular cases, when either $a=0$ or $b = 0$, the number of different
elements in Eq.(\ref{Wlambda}) is reduced from 6 to
3 (see Fig.~1a), and it is only 1 if $a=b=0$. Note that the 
action of the Weyl group
expands the fundamental region {\sf F} onto the maximal torus {\sf T}.
  
An {\it orbit function} $\Psi_{(\lambda) } \equiv \Psi_{a,b} $ at the vector 
point $r$ (hereafter we use this simple notation instead of {\bf r}) 
is defined ~\cite{MP1,MP2} as a finite sum of exponential functions on 
the Weyl group  orbit of the element $\lambda$:  
\begin{equation}\label{Psi_lambda}
\Psi_{a,b} (r) = 
\sum_{\mu \in W(\lambda)} e^{2 \pi i (\mu , r )}, 
\end{equation}
where $(\mu , r)$ is the scalar product of 2 vectors. Here we will  
be interested in orbit functions corresponding to integer indices $a $ 
and $b$. 

 An EFO is an element $X$ of the group, such 
that $X^N$  is the unit element $E$ for some natural number N. 
In the $\omega$-basis representation $X = a_1 \omega_1 + b_1 \omega_2$ this 
condition is satisfied if (and only if) $a_1$ and $b_1$ are rational numbers. 
The sets of all EFOs of adjoint order dividing $N$
are represented by vector points $X\rightarrow {r}_{k,m}$ 
with $a_1= k/N$ and $b_1 = m/N$ that satisfy
the conditions $0 \leq k, m, k+m \leq N$. These make sets $\{ r_{km}
\} \equiv {\sf F}_N \subset {\sf F}$ in the form of equilateral 
triangular grids, as shown in Fig.~1b for $N=12$. 

The key property for the method of DFT  
on compact semisimple Lie groups   
is the discrete orthogonality of orbit functions with 
different $\lambda$ on the sets ${\sf F}_N$ of EFOs of adjoint order $N$ 
~\cite{MP1,MP2}.  
In case of SU(3) group this property corresponds to the equation 
\begin{equation}\label{orthogonality}
\sum_{r_{km}\in {\sf F}_N} P_{km} \Psi_{j,n}(r_{km}) 
\overline{\Psi_{i,p}}(r_{km}) =  \delta_{i,j} \delta_{p,n} \, 
D_{N}(j,n)
\end{equation} 
where 
\begin{equation} \label{multi}
P_{km} = \frac{6}{(1+\delta_{k+m,0}) (1+\delta_{k,0} +\delta_{m,0})} \; ,
\end{equation}       
is a multiplicity factor\footnote{it shows the number of 
elements in the torus {\sf T} which are conjugate to the given 
$r_{km} \in {\sf F}$},
and the Kroneker symbols are treated modulo N,  i.e. 
$\delta_{i,j}= 1$ if $i = j$ mod($N$), e.g. $\delta_{N,0} = 1$. 
The normalization factor in Eq.(\ref{orthogonality}) is  
\begin{equation}\label{DN}
D_{N}(j,n) =  \frac{108 N^2}{P_{jn}} \; , 
\end{equation} 

Important statements for the method of DFT on Lie groups ~\cite{MP1,MP2}
 are the following two: \\
({\bf a})  for any given $N$ the set of orbit functions $\Psi_{j,n}$ 
with indices $0 \leq j, n, j+n \leq N$ makes  
a full set of functions orthogonal to each other on the equilateral triangular
grid $\{r_{km} \mid 0 \leq k, m, k+m \leq N\}$ in the form of 
Eq.(\ref{orthogonality}); \\
({\bf b}) this set provides a basis for DGT on SU(3) orbit functions 
with the smallest possible wave numbers $(j,n)$.

A DGT of a given discrete function $\{ G_{km}\equiv G(r_{km}) \}$
produced by sampling of a continuous ({\it analog})
function $G(r)$ at the grid  points $\{ r_{km}\}$  
corresponds to  solution of the set of equations
\begin{equation} \label{invDGT}
G_{km}= \sum_{j,n \geq 0}^{j+n \leq N} A_{jn} \Psi_{jn}(r_{km}) \; \;  
{\rm for} \; \; 0 \leq k,m, k+m \leq N \, .
\end{equation}
Multiplying Eq.(\ref{invDGT}) by $P_{km} \overline{\Psi_{ps}}(r_{km})$,  
then summing up  over $\{ k, m\}$, and using Eq.(\ref{orthogonality}), 
the transform values $\{ A_{jn}\}$  are readily found:  
\begin{equation} \label{DGT}
A_{jn} = \frac{1}{D_N(j,n)} \sum_{k, m \geq 0}^{k+m \leq N} P_{km}
G_{km} \overline{\Psi_{jn}}(r_{km}). 
\end{equation}

\subsection{Continuous extension of DGT}

The set $\{ A_{km} \,\mid \, 0\leq k, m , k+m \leq N  \}$ is 
an exact Fourier transform of $\{ G_{km} \} $,  
as far as the knowledge of $\{ A_{km} \}$ allows unambiguous  
reconstruction of $\{ G_{km}\}$ at the points $\{ r_{km} \}$ 
of the given grid using the {\it inverse} DGT, i.e. Eq.(\ref{invDGT}). 

The inverse DGT
can be extended into a continuous function in the form 
of Fourier polynomial (i.e. a series of cosine and sine 
functions) as suggested in ~\cite{CEDCT}. It is done 
simply by replacing the discrete spatial argument $r_{km}$ of the orbit 
functions $ \Psi_{jn}(r_{km})$ in the inverse DGT series, 
Eq.(\ref{invDGT}), by  the continuous argument $r$. 
The resulting function 
\begin{equation} \label{CEDGT}
F_{N}(r) =   \sum_{j,n \geq 0}^{j+n \leq N} A_{jn} \Psi_{jn}(r)
\end{equation}
is called {\it continuous extension} of DGT,
or CEDGT. Note that in case of real functions, $F_{N}(r)$ is reduced to
$\sum_{\{j,n\}} [\, {\rm Re} A_{jn} \, {\rm Re} \Psi_{jn}(r) - {\rm Im} A_{jn} 
\, {\rm Im} \Psi_{jn}(r)\,]$.

The function $F_{N}(r)$ coincides, obviously, 
with $\{ G_{km}\}\equiv G(r_{km})$ at the 
grid points. However, the most valuable property of $F_{N}(r)$
is in the quality of its interpolation of the values $G_{km}$ 
between the grid points. The approximation to the original $G(r)$
that it provides is such that even the first
and (under certain conditions) also the second derivatives of 
$F_{N}(r)$ are meaningful functions, converging to the respective
derivatives of $G(r)$. The CEDGT also 
satisfies the property of {\it locality} of the transform. These
are similar to the properties of 
canonical {\it continuous} Fourier transforms. Note that  
these properties do not hold, however, for the continuous extension 
of the traditional version of DFT (i.e. which is often used in case of
rectangular grids). Even the straightforward continuous extension of 
the ordinary DFT is a meaningless (profoundly {\it oscillating}) function   
(see ~\cite{CEDCT} for details).

For convenience here we write the explicit expressions for the
SU(3) orbit function $\Psi_{a,b}(r)$. 
In Cartessian coordinates  with the $y$-axis   
bisecting the equilateral triangle (i.e. the fundamental region), 
with the sides of length 1,  as in Fig.~1b, 
the real and imaginary parts of the orbit functions 
are reduced to the following:
\begin{eqnarray} \label{Psi}
{\rm Re}\, \Psi_{a,b}(x,y) & = & 
\; \; 2 \, \cos\left(2 \pi y \frac{a+b}{\sqrt{3}}\right)
\cos\left(2 \pi x \frac{a-b}{3}\right)  \nonumber \\
& & + \: 2 \, \cos\left(2 \pi y \frac{a}{\sqrt{3}}\right)
\cos\left(2 \pi x \frac{2b +a}{3}\right) \\
& & + \: 2 \, \cos\left(2 \pi y \frac{b}{\sqrt{3}}\right)
\cos\left(2 \pi x \frac{2a +b}{3}\right) \; , \nonumber \\
{\rm Im}\, \Psi_{a,b}(x,y) & = & 
\; \; 2 \, \cos\left(2 \pi y \frac{a+b}{\sqrt{3}}\right) 
\sin\left(2 \pi x \frac{a-b}{3}\right)  \nonumber \\
& & + \: 2 \, \cos\left(2 \pi y \frac{a}{\sqrt{3}}\right)
\sin \left(2 \pi x \frac{2b +a}{3}\right) \\
& & - \: 2 \, \cos\left(2 \pi y \frac{b}{\sqrt{3}}\right)
\sin\left(2 \pi x \frac{2a +b}{3}\right) \; . \nonumber 
\end{eqnarray}
In this system of coordinates, the property of  
complex conjugation of orbit functions  
$\Psi_{a,b} = \overline{\Psi_{b,a}}$ results in useful  
symmetry properties of the real and imaginary parts 
with respect to variables $x$ and $y$. Also, if $a=b$, then
${\rm Im}\, \Psi_{a,a}(x,y) =0$. Note that 
in case of integer $a$ and $b$ the harmonic 
order of these polynomials is not always integer.

\subsection{Examples of CEDGT for analytic functions sampled on grids}
  
The interpolation power of continuous extensions of DGTs is 
demonstrated in Fig.~2. For an example, here we interpolate a function
sampled on a rectangular grid. The appropriate Lie group for this symmetry 
is SU(2)$\times$SU(2).
The DGT in this case happens to result in  
2-dimensional case of a 
transform known as DCT ({\it discrete cosine transform}). The orbit functions of 
the SU(2) group, for a one-dimensional function, are composed only of 2 
exponents, 
$\Psi_\lambda (\theta)
=e^{i 2 \pi \,\lambda \theta} + e^{- i 2 \pi \,\lambda \theta}
=2 \cos (2\pi \lambda \theta)$ with
$\theta \in [0,1/2]$ ~\cite{CEDCT}. The basis for the two-dimensional DCT 
simply is a product of 2 cosine functions in 2 orthogonal 
directions in the plane.

The discrete image shown in Fig.~2a 
is produced from sampling of the original analytic function $G(r)$  
composed of a sum of two
2-dimensional Gaussian distributions with large axes inclined at an angle
25$^\circ$ to each other.  The characteristic widths of both 
are chosen to be smaller than the distance between the grid points, 
and correspond to dispersions $\sigma_\perp = 2/3N$. 
This grid is rather sparse
for the image, which is apparent on Fig.~2a. 
The result of application of CEDGT is demonstrated in Fig.~2b 
(panel on the right). The 
original function is recovered almost exactly, clearly separating the 2 
ellipsoids and recovering their orientations. The only noticeable difference
between CEDGT function $F_N(r)$ and the original $G(r)$ is reduced to some 
low-amplitude wiggles.  In Fig.~2b these wiggles result in the appearance 
of the dashed contour line corresponding to the level of 
$-0.1\%$ of the maximum intensity in the image.

It should be noted that a straightforward interpolation of discrete 
images by CEDGT can be useful in case of low level of
noise in the signal, but it may be less effective if the level of
noise is high. This is because the CEDGT `zooming' of an image 
containing random noise makes the latter more apparent at small 
spatial scales. 
In this regard, an important property is that the Fourier transform of  
an additive random (`white') noise also is 
a random noise, whereas the DGT representation $\{A_{jn}\}$
of any meaningful image is concentrated in the domain of long frequency
harmonics (e.g. see ~\cite{DCT}). That is, the coefficients 
$\mid A_{jn}\}\mid$
have a general tendency to decrease with the increase of $j$ and $n$. 
Therefore the noise typically becomes relatively stronger, and even may 
completely dominate the signal at high frequencies, as demonstrated in Fig.~3. 
The big dots here correspond to the mean absolute values
$\mid A_{km} \mid$ of the transform coefficients of different harmonic
orders $K=k+m \leq N$ for an image presented in Fig.~4a.     
The stars show the mean values of the DGT coefficients of the noise
(contributed by 2 'hot pixels' described below in Fig.~4) in this image 
alone. 
It is obvious that at high frequencies  $K\geq N/2$ the 
the DGT image is totally dominated by this noise. 

This property can be exploited for organizing a very simple, but also very 
effective, low-pass filter. 
Namely, we can cut off in the frequency domain all modes $A_{km}$ 
of harmonic orders exceeding some large $N_{cut}$, i.e. putting 
$A_{km} \rightarrow 0$ for indices $ N_{cut} < K=(k+m) \leq N$. Then we can 
use the truncated CEDGT series that contains only the modes $K\leq N_{cut}$
to approximate the image. This filter can be described by a parameter 
$C_f = 1 - N_{cut}/N$
such that $C_f=0$ corresponds to "no filtering".

The result of application of such low-pass filter is shown in Fig.~4.  
The original image in Fig.~4a  
corresponds to  continuous function in the form of 2-dimensional Gaussian 
distribution with effective width $\sigma_\perp = 1/N = 0.05$ and 
length $\sigma_{\parallel} = 2\sigma_\perp$.
Onto it  2 isolated `hot pixels', with heights of each
equal to the 
half of the maximum of the main image, are superimposed. One of the 
'hot pixels' is chosen far from the main structure, but the second 
one falls within the image close to its maximum. This results in the total 
signal in that pixel exceeding the maximum value of the 
main function.

Fig.~4b (panels in the middle) demonstrates the result of
straightforward extension of the discrete image using CEDGT without
filtering. Note that the amplitude of oscillations of the
hexagonal Fourier waves in the near vicinity of the isolated `hot pixel' 
may reach up to $\simeq 15$-$20\,\%$ of the 
`hot pixel' value. 
The distance between the contour lines corresponds to $5\,\%$ of the 
maximum of the ellipsoid, and dashed contours correspond to the negative 
values of the CEDGT image. Note that the amplitude of the  
oscillations rapidly declines with distance from the `hot pixel' 
because of the {\it localization} property of the CEDGT ~\cite{CEDCT}.

In Fig.~4c we show the result of application of the low-pass 
filter with the filtering parameter $C_{f}= 0.5$. 
The main signal practically does not change. This is because 
for the original images with effective widths $\sigma_\perp \geq 1/N$, 
the Fourier transformed `image' $\{A_{km}\}$  is concentrated
mostly in the domain of low wave numbers $(k+m) \leq N/2$, as shown in
Fig.~3. Note that even for more narrow images, with widths about only 
one half of the size of pixels in the grid,
 $\sigma_\perp \simeq 1/2N$, the amplitude of the image after filtering
 with $C_f=0.5$ drops only by 
$\sim 25$-$30\,\%$, but the orientation 
of the original images is recovered quite well.     
Meanwhile, both 
"noisy pixels" have almost disappeared from the filtered 
image in the Figure 4c. 
The signal in the isolated pixel has dropped by a factor $\sim 4$. 
Even more important is that the intensity in 
the `hot pixel' inside the main body of the image has practically recovered 
the expected value of the signal. The difference between the 
contours in Figures 4a and 4c is mostly at contour levels
corresponding to low intensities. This difference 
can be completely eliminated by modest `tail' cuts,
which is the standard image pre-processing procedure discarding the 
pixels with signals below some level, 
or perhaps even more effectively, by the `height' cuts as we explain below
in Section 3.

\section{Performance of the FT method}

To compare relative performances of the FT- and S- methods,
Monte-Carlo data bank for proton and gamma-ray primaries has been simulated 
using Hillas' MOCCA code ~\cite{MOCCA} complemented with the full ray-tracing 
of the incident Cherenkov photons ~\cite{rtracing}.
The IACT parameters, such as photocollector radius/size, altitude, etc.,
have been assumed corresponding to the telescopes of the HEGRA
system located at $2200$ m above sea level (see ~\cite{HEGRA}). 
In particular, the
images are formed in the plane of photoreceiver  consisting of
271 PMTs (or {\it pixels}), with the angular size $h=0.25^\circ$
 arranged in the hexagonal grid. This grid is relatively coarse
compared to the cameras of the contemporary IACT projects, and also the number
of PMTs is relatively modest. Both these factors do help, however, to compare
and reveal more easily the differences between the FT- and S- methods.

For  both proton and gamma-ray events, the showers have been generated 
with energies starting from $0.3\rm TeV$, which is significantly
lower than the detection threshold of IACTs with the given parameters. 
The simulations were performed 
for a telescope pointed to the zenith. The gamma-ray EAS with power-law 
index of 
the primary photon spectrum $-2.6$ are generated in the vertical direction,  
i.e. parallel to the telescope 
optical axis. For the CR protons, an isotropic distribution within a
vertical cone with the opening (half-) angle  
$\leq 2.7^\circ$, and  with the spectral index  
$-2.67$ has been used. The trigger condition 
for EAS detection was chosen as `2nn/217 $\geq 10\,pe$', i.e. $\geq 10$
 photoelectrons
in any 2 near neighbors from 217 inner PMTs. This excludes 
the PMTs of the last (hexagonal) ring  from the trigger. 
In our data bank the number of proton EAS  passing the trigger  
is $N_{p, tot}=2354$, and $N_{\gamma, tot}=4229$ for gamma-ray events.

\subsection{CEDGT representation of the EAS images} 

In order to apply the DGT, we have to embed the hexagonal grid 
of the camera into the equilateral 
triangular grid, i.e. into the fundamental domain of SU(3), 
as it is shown in Fig.~5. 
The camera with 271  
pixels contains $M=9$ `hexagonal rings' of PMTs around the central PMT. 
We formally add one more
ring assuming $S_{km}=0$ for the numbers of photoelectrons there. It gives
some additional space for processing images that may contain non-zero
numbers of photoelectrons in the 9-th ring of the camera. All this results 
in a grid ${\sf F}_N$ having $N=30$ equal intervals, and 
$N+1=31$ grid points, along the 2 principal {\it non}-orthogonal 
axes  $\omega_1$ and $\omega_2$  of the triangle. 

Our next step is to reconstruct the values $G_{km}$ for the photoelectron
 distribution (the Cherenkov light brightness) function $G(r)$  at the 
grid points $r_{km}$. This is a useful pre-processing procedure, 
as far as the numbers $S_{km}$ in the detected
image represent the values of photoelectrons 
{\it integrated} over the surface of 
each individual PMT. Therefore  $S_{km}$  represents only the mean
of the continuous distribution function over the PMT, and only in the 
0-th order approximation it represents the real 
brightness distribution $G(r_{km})\equiv G_{km} \simeq S_{km}/s_{\rm pix}$
in the centres of PMTs.  
Here $s_{\rm pix}$ is the PMT surface in square angular units, which
can be chosen as a normalization factor, i.e. $s_{\rm pix}=1$. 

The values $G_{km}$ are better recovered if we assume that 
$G(r)$ is
a continuous function that can be approximated within each PMT
by 2-dimensional Taylor series around its centre $r_{km}$. 
Integrating over the hexagonal surface of the PMT, 
we arrive at an expression that connects the set $\{G_{km}\}$ with 
$\{ S_{km}\}$ through the 
second-order derivatives of $G(r)$ at the grid point $r_{km}$:
\begin{equation}\label{differential}
G_{km}\approx \frac{S_{km}}{s_{\rm pix}} 
- \frac{5 h^2 }{54} \, [ G^{\prime\prime}_{x}(r_{km}) +
 G^{\prime\prime}_{x+60}(r_{km}) + 
 G^{\prime\prime}_{x-60}(r_{km} ) ] \; .
\end{equation}
Here $G^{\prime\prime}_x$,  $G^{\prime\prime}_{x+60}(r_{km})$ and
 $G^{\prime\prime}_{x-60}(r_{km})$ are the second derivatives of $G(r)$
along the 
$x$-axis and along the directions of fundamental weights $\omega_1$ and 
$\omega_2$ of the equilateral triangular grid, respectively,  
i.e. at the angles $\pm 60^\circ$ relative to the $x$-axis (see Fig.~1b).
 This equation allows fast 
 iterations, approximating initially the derivatives 
of the 
grid functions in the standard difference scheme approach as 
$f^{\prime\prime}(r_0)
= [f(r_0 +h) + f(r_0 -h) -2 f(r_0) ]/h^2$. Here the notations $+h$ and $-h$ 
indicate the  positive and negative shifts, respectively, in the 
given direction along the grid from the grid point $r_0$. 
For the subsequent iterations the derivatives 
are calculated using directly the CEDGT function.  
This approach provides a good agreement between the 
integrals of the brightness distribution function and the total 
numbers $\{S_{km}\}$ in the PMTs already after the second iteration. 
The agreement better than 
$\leq 1\,\%$ for most pixels, except for some (but not all)
of those with very low numbers $S_{km}$. The total number of photoelectrons 
in the image is preserved within $\leq 1$-2\,\% accuracy.

In Fig.~5 we show the images of a $E=1\,\rm TeV$ gamma-ray EAS incident at
a distance 130m from the telescope. Fig.5a (panel on the left) 
shows the raw image detected by the camera, and Fig.5b (panel in the 
middle) shows the continuous distribution function provided by CEDGT.
Note that the lowest contour shown corresponds to the level of 3\% from
the maximum value of the image. It demonstrates that the ripples of the
Fourier waves are very significantly eliminated already at this very low
cutoff level\footnote{Note that for our M-C data bank the level 
``$3\%$ from the image maximum'' corresponds on average to 1 photoelectron.}.  
In Figure~5c the image is reconstructed after application of the 
low-pass filter with parameter $C_{f}=0.35$. The image here is 
unified into an essentially single-core pattern, as expected for the 
gamma-ray events. The contour levels in Figure~5c are at the same 
absolute values as in  Figure~5b. Comparing with the contours  
in Figure~5b shows a drop in the amplitude of the maximum by $\sim 40\%$.
This is because in this particular image the signal in one of the 
pixels was much higher than in all of its nearest neighbors. The filter
accepts that as a (statistical) fluctuation and smooths it out. The total 
number of photoelectrons in this image makes in fact $\simeq 85\%$
of the original $S_{tot}$. 

Figure 6 shows the results of application of the same  
filter with $C_f=0.35$  to the image of a proton EAS with energy $E=2\,\rm TeV$
falling at a distance $100\, m$ from the telescope.
The maximum of the filtered distribution in Figure~6c 
is about $85\,\%$ of the unfiltered CEDGT image, and the total number of 
photoelectrons found after integration of the filtered image is
about $90\%$ of the original $S_{tot}$. This implies that the 
effective width of the filtered image becomes somewhat larger
than of the original. The most important feature is, however, that
unlike the gamma-ray image, the proton image shows the main EAS core
but it is still far from unification into a single-core pattern. 
Two distinct `islands', presumably connected with the EAS pions, 
are still apparent.

\subsection{Application to `pure' EAS images}

The efficiency of the FT-method for signal enhancement is compared with 
the S-method using 
 the orientation parameter ALPHA. This parameter is  
chosen since it is known as one of the most effective parameters 
(along with AZWIDTH) currently  
used for the IACT image processing in the standard parameterization scheme. 
It is also known that ALPHA corresponds to 
the deflection angle of the image major axes from the direction
to the central pixel of the camera viewed from the image centre of mass 
(i.e. from the presumed source direction, see e.g. ~\cite{ALPHA}). 
The signal enhancements provided by FT- and S- methods at different 
levels of tail-cuts $c$ 
and ALPHA-'cuts' (i.e. chosing ALPHA$\leq \alpha$), 
are compared in terms of Q-factors, 
$Q\equiv Q(c,\alpha) = \eta_{\gamma}/\sqrt{\eta_p}$. Here    
\begin{eqnarray}\label{eta}
  \eta_{\gamma}\equiv \eta_\gamma(c,\alpha) & = &
\frac{N_{\gamma}(c,\alpha)}{N_{\gamma , tot}} \; ,\\
\eta_{p}\equiv \eta_p(c,\alpha) &  = & \frac{N_{p}(c,\alpha)}
{N_{p,tot}} \; \nonumber
\end{eqnarray}
are the fractions of gamma-ray and proton EAS, respectively, which 
remain in the pool after corresponding parameter cuts $c$ and $\alpha$. 
The maximum values $Q_{\rm max}$ are calculated by applying first a fixed 
tail-cut, and then varying parameter $\alpha$. Note that we consider only
$Q$-factors which preserve at least $50\%$ of the initial gamma-ray events 
$N_{\gamma,tot}$, i.e.
$\eta_{\gamma} \geq 0.5$.    

To calculate the FT-image, we first reconstruct the 
values $\{G_p\}$, using Eq.(\ref{differential}), for the continuous 
distribution function of photoelectrons in the centres of PMTs that 
would correspond to the total numbers
of photoelectrons $\{S_p\}$ in the image. Here the subscript  
$p=1,2,...P$ stands for the grid coordinates $(i,j)$ of each of $P$ 
pixels involved in the  
image. The Fourier transform coefficients $A_{i,j}$ are calculated,
and the coefficients which do not pass the given filter $C_f$ are discarded
from the CEDGT series of Eq.(\ref{CEDGT}). Using this truncated 
CEDGT function, the 
values of the brightness distribution $\{ G_{k} \mid k=1,2,..., K \}$ 
with $K \gg P$ can be calculated for a new grid with any higher  
density of points. 
For our calculations we divide each of the 6 equilateral triangles 
of the hexagon (the PMT surface) into 4 equal sub-triangles, and 
calculate the values $G_k$ in 
the centers
of each of these sub-triangles. Thus, the density of points in the new 
image is increased by a factor 24. This also leads to an increase of the  
total number of points in the image approximately (because of some zeros) 
by the same factor, i.e. 
$K\simeq 24\,P$. After that a 
"tail cut" (or "height cut", see Section 3.3) procedure is applied, i.e.
$G_{k} \rightarrow 0$ if $G_{k} < g_{cut}$

Note that here we implement a slightly modified than the currently 
used tail cut procedure.
Namely, the tail cut level $g_{cut}$ is not fixed to some 
value that is the same for all images, as in the standard
procedure.  Instead, for each image 
we use the "percentage cut", where the cut level 
$g_{cut} = c \times G_{max}$ is defined as 
a fraction $0 \leq c < 1$ from the image maximum $G_{max}$  
to cut, but where this fraction is now fixed.  This procedure takes better 
into account that the amplitude of possible wiggles in the CEDGT function
would linearly increase with $G_{max}$. In order to make comparing with the
S-method more unambiguous, we also use the same "percentage cut'' 
approach for the S-method. We have checked that the maximal 
Q-factors reached in the S-method in case of both of these tail-cut schemes   
are practically the same. Note also that in order to reduce the  
impact of statistical fluctuations due to limited 
numbers of events in the $(c_i, c_i+\Delta c;\, \alpha_j, \alpha_j+
\Delta\alpha)$-bins, we first apply a standard
technique of averaging of the numbers of events in the bins in the 
$(c,\alpha)$ plane
over $(3\times 3)$ square window of bins centered at the given $(i,j)$-bin.
Only then the Q-factors are calculated. This procedure makes all the results
more systematic, and therefore more conclusive.

In Figure~7a we present the results of application of the FT-method
to our M-C data bank. Here we show the maximal values
of the $Q$-factors reached at different tail-cuts when the parameter
$\alpha$ is allowed to vary.
The curves 1 and 2 show the maximal $Q$-factors 
for the S-method and for the FT-method without filtering 
($C_f=0$), respectively. Both provide practically
the same $Q_{max} \approx 2.92$, although at somewhat different
tail-cuts. The curve 3
shows the maximal Q-factors when the filter
with $C_f= 0.45$ is applied. The absolute maximum of that curve is 
significantly increased, reaching $Q_{max} \approx 3.4$
In Fig.~7b we show the $\alpha$-behaviours
of the Q-factors for the same 3 cases, but when the tail-cuts 
are fixed at those ({\it different}) values that produce the 
respective maximal signal enhancements. Systematic gain in the 
signal enhancement after image filtering is apparent in both 
Figs. 7a and 7b, which   
proves that the filter does work.

Remaining in the framework of image discriminators based on 
standard parameterization, it seems reasonable to propose that the FT-method
would be relatively more effective than the S-method 
for relatively poor images, when the numbers of the pixels involved
become small, $P\sim 10$ and less. This is because for `rich'
images containing large numbers of active pixels, the calculated 
moments of a discrete image should be less affected by the sparsity 
of the data than in case of a coarser image. A photon-`rich' image is
sufficiently `smooth' already, therefore an additional smoothing 
of the image by CEDGT would not result in a significantly better 
reconstruction of the image orientation. It would not be so, however, for 
the photon-poor images.  

The number of active pixels $P$ in the image is correlated 
with the total number of detected photoelectrons $S_{tot}$.  
Therefore in order to check the validity of that proposition, 
we divide the data bank into 2 subsets containing images with 
$S_{\rm tot} \leq 200\,$pe and $S_{\rm tot} > 200\,$pe. 
The value 200 is chosen from the consideration to have 2 subsets containing 
statistically significant numbers of both gamma-ray and proton events in 
both "photon-poor" and "photon-rich" data banks. 
These numbers are equal to $N_{1,p}=1321$
and $N_{1,\gamma}=2843$ for the photon-poor set, and $N_{2,p}=1033$
and $N_{2,\gamma}=1386$ for the photon-rich set. 

In Figure 8  we compare the behaviours of maximal $Q$-factors at different 
tail cuts for these 2 subsets separately. For both methods the maximum 
signal enhancement  has been increased in case of photon-rich images, 
 reaching $Q_{max} \approx 3.8$ for both. At the same time,
for photon-poor images the difference between the standard and FT
approach is increased further. It is important that for the photon-poor images
the deterioration of the maximum $Q$-factor in the FT-method is relatively 
small, by only $\simeq 0.2$, so $Q_{max} = 3.24$.
Meanwhile the maximum Q-factor in the S-method
declines down to $Q_{max} = 2.64$. 

 These results suggest that the Fourier transform method allows
to work with photon poor images significantly better than S-method 
does. The signal could be distributed only in few pixels, but the 
orientation of the gamma-ray EAS could be still recovered with
sufficient accuracy by the FT-method. Note that even in hypothetical 
case of only 2 pixels involved in the image, the method attributes
a minimum dispersion to it, which is of the order of one half of the 
pixel size. In practice, one can process images containing only 
$\geq 3$ pixels.
FT-method allows to perform shape analyses involving any deep cuts 
 even for these very sparse images.  

\subsection{Processing of noisy images}

Photon-poor images are produced mostly by the events with
energies of incident particles closer to the detection threshold of 
the telescope. 
Thus, the results shown in Figures~8a and 8b seem to suggest 
that the FT-method contains a significant potential to deal  
with the low-energy events better than the S-method. This implies
a potential for substantial reduction of the energy thresholds of 
the detectors. The question by how  
much exactly the IACT energy thresholds could be reduced with the  
use of the proposed here FT-method requires a separate detailed 
study out of the scope of this paper. 

Here we only note that in practice the number of low-energy events detected 
strongly depends on the hardware trigger condition used. Relaxing the trigger
condition in order to reduce the detection threshold 
in real experiments results in a very significant increase of the number of
events with low signal-to-noise ratio. 
In this regard, an important question is whether the FT-method would
also allow to deal with images with enhanced level of noise over the 
entire camera that would be copiously detected in case of softening 
 the hardware trigger condition.

In order to address this question, 
we simulate random `white' noise over the entire camera with the mean 
number of photoelectrons per pixel $\bar{n}_{0} = 2$. 
Poisson distribution
with this mean $\bar{n}_{0}$ has been used to generate the noise
$n_{p}$ in each of the 271 individual pixels. This noise has been 
added to the M-C  
images of EAS, resulting in  
$S_{i}^\prime = S_{i} + n_{i}$. The trigger condition has been applied 
before adding the noise, in order to have the 
same EAS images in the  `pure' and `noisy' sets. 

For noisy images it can make sense to use `height cuts', instead 
of usual `tail cuts'. Using again the fixed percentage cuts 
$g_{cut} = c G_{max}$
as in the tail cuts, the `height' cuts correspond to  
image modification procedure $\tilde{G}_{p} = \max (G_{p}-g_{cut}, 0)$.
After such `height' cuts the image momenta should be less sensitive 
to the noise in distant pixels than after tail cuts which do not modify the
signal amplitudes in the pixels passing the cut. 
Therefore one could expect a faster, i.e. at lower cut values $c$, 
noise suppression and recovery of the  
signal at height-cuts than at tail-cuts. Note that one 
should not expect a full recovery of 
the initial $Q$-factors, because also the pixels containing the signal are 
affected by the noise. 

The recovery of the Q-factors for noisy images by S- and FT-methods
is shown in Figures 9a and 9b using height cuts and tail cuts, respectively.
For comparison, in both panels we also show the Q-factors for the total
set of pure
(i.e. initial) images without noise. Comparing in these 2 panels the 
curves with full and open squares for the FT- and S-methods, respectively,
 we can see that both types of cuts result practically in the
same best values for $Q_{max}$ in each of these methods if the  
images are `noise-free'.
All other curves are for the noisy data
from which the mean value $\bar{n}_0$ of the simulated noise is uniformly
substructed from each pixel, i.e.  $S_{i}^{(2)}=\max(S_{i}^\prime - 2; 0)$.
Note that this is similar to 
substruction of a `pedestal' of $2pe$. Both panels demonstrate that
the signal enhancement capability of the S-method and the FT-method 
without filtering ($C_f=0$, curves 1) 
has been essentially killed by noise.
One needs very strong height or tail cuts at the level of $30\%$
in order to bring the Q-factors back at least to the level of $\simeq 2$. 
Meanwhile, after cutting off the high frequencies the FT-method
is able to recover the Q-factors to the level of $2.7$.
One can also see from comparing the curves marked `2' on both panels 
that the height cuts do provide a faster recovery of the Q-factors
than the tail cuts, although the final results at very large cuts
are similar (as expected).

In Figs. 10a and 10b we show the rates of recovery of Q-factors
when different levels of `pedestal substruction' are
applied prior to applying the height cuts. 
Curves marked by numbers $k=2,4,6$ correspond to noisy
data from which $k$ photoelectrons are substructed, i.e. for each 
pixel we take the 
maximum $S_{i}^{(k)}=\max(S_{i}^\prime - k; 0)$. 
For the FT-method we use the filter $C_f=0.45$. The maximal Q-factors 
reached in each of these 3 cases by the FT-method 
are practically the same. The only difference between these cases 
consists in faster recovery of the signal when larger $k$ is substructed,
as one expects. Meanwhile, with the standard
method the maximum attainable Q-factors are significantly lower
than with the FT-method, and also are rather sensitive to the 
amount of pedestal substruction applied. 

Fast recovery of the Q-factor using FT method for image processing 
makes possible extraction 
of the signal at lower cut levels. It allows to keep more images in the
raw data, in the sense that photon-`poor' images can be processed more 
effectively, and using height or tail image cuts at much deeper levels than
the standard method would allow.

\section{Discussion}

The technique of discrete Fourier transform on orbit functions of
Lie group SU(3) 
allows processing of discrete data/images given on the uniform
grids of triangular or hexagonal symmetry. This is the symmetry implemented
in all of the contemporary IACTs, including
HESS ~\cite{hess}, MAGIC ~\cite{MAGIC}, VERITAS ~\cite{veritas} and 
CANGAROO-III ~\cite{cangaroo3}. 
In case of rectangular PMT grids, such as used in 
CANGAROO-II ~\cite{cangaroo2}, 
the appropriate DGT for implementation of the FT approach proposed 
in this paper corresponds to the group SU(2) ~\cite{CEDCT}. 

The DGTs allow straightforward {\it continuous extensions}, in the form 
of trigonometric polynomials, from the discrete
points of the grid to any point on the image plane. 
The CEDGT functions $F_N(r)$ are distinguished by a 
number of valuable properties, including {\bf (a)} 
{\it convergence} to the original function $G(r)$ with increasing $N$; 
{\bf (b)} {\it localization} property; and {\bf (c)} {\it differentiability} 
of $F_N(r)$, implying convergence of
the derivative of  $F_N(r)$ to $G(r)$ ~\cite{CEDCT}.
These properties are very similar to the properties of the canonical 
Fourier transforms of {\it continuous} functions.

The analysis on the M-C simulated data bank presented here
shows that CEDGT approach 
can be effective for enhancement of gamma-ray signals. 
Remaining in the framework of standard parameterization, 
i.e. without exploring new possibilities that can be opened with 
the use of FT approach, the Q-factor for the parameter ALPHA 
is steadily enhanced from $Q_{max} approx 2.9$ in the S-method 
to $Q_{max} = 3.4$
simply by using the CEDGT with the low-pass filter
$C_{f} =0.45$. The latter is a simple possibility provided by
the Fourier method to suppress the statistical fluctuations or noise
(of different origins) from the data.
Our calculations indicate that the useful range for the
filter parameter is within $C_{f} \simeq (0.3-0.5)$, depending on the 
type of image analysis which is being carried out. 
For the orientational
parameter ALPHA values closer to 0.5 seem to be optimal, and in this paper
we have fixed it to $C_{f}=0.45$. However, one might also 
chose values $\sim 0.3$
or even less for studies where a lesser degree of broadening and unification
of proton images would be advantageous. 

Compared with the standard method, the power of continuous extension 
of DGTs for image processing becomes more apparent for photon-poor
images, as demonstrated in Figs.~8a and 8b. Furthermore, the FT-method 
offers to deal much better that S-method with images 
where the level of random noise over camera is substantially
enhanced. The implication is that perhaps with the use of FT-method 
the effective energy threshold of EAS detected by IACTs could be 
further decreased. A more definite 
answer to this question requires, obviously,  separate studies 
involving relaxed trigger conditions, and possibly also considering lower 
 than currently levels of the pedestal substruction. 

Out of the scope of this paper also remain possibilities of the FT-method 
for the stereoscopic systems of IACTs. 
The single notice in this respect is that 
the power of FT-method for better reconstruction of the EAS direction and 
keeping more gamma-ray events in the ``single telescope pool''
should obviously show up in the systems of IACTs as well (and may only be 
amplified).  

Neither we consider here any new possibilities for 
image processing that might be possible with this technique.
Those methods can be connected either with image processing in the
image domain itself (such as image analyses under deep
 `tail cuts' and `frequency cuts', use of image derivatives, etc),
or image analyses in the frequency domain (analysis and optimization of the 
FT coefficients).

At last, we would like to note that the technique of Fourier transforms
described here is unique for grids with hexagonal symmetry, and it 
is easy to use in practical calculations. Therefore it might be 
useful for data processing not only for the IACTs, but also 
for other detectors using hexagonal symmetry, 
such as HIRES detector in ultra-high energy CR physics,  
or the IceCube detector in VHE neutrino astronomy.

\begin{ack}
The authors appreciate the support of this research by the NATO Collaborative 
Research Grant, Ref. PST.GLS.979437, which made this work possible. 
One of us (J.P.) acknowledges support from the National Science and
Engineering Council of Canada (NSERC) through the Discovery
Grant programm. We are also greatful for partial support
for this work from Lockheed Martin Canada and NSERC through
the MITACS, Network of Excellence for Mathematical Sciences.
\end{ack}

\clearpage

\begin{figure}
\centerline{
\epsfysize=6.cm \epsfbox{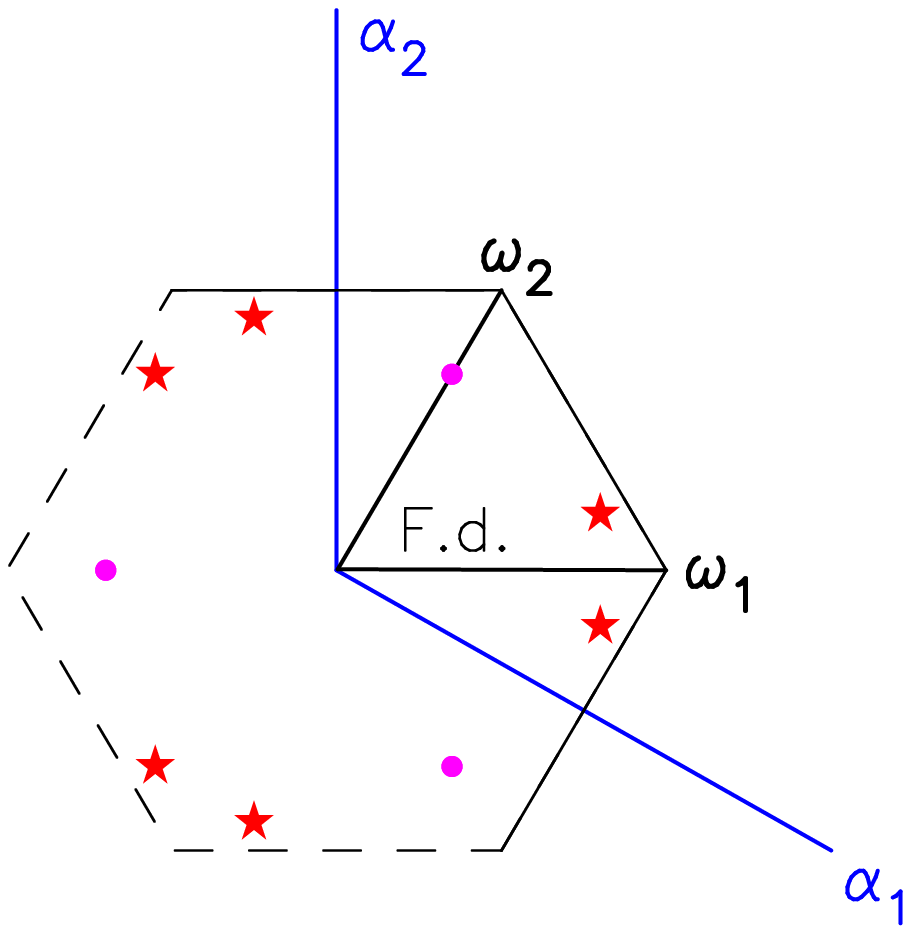}\hspace{5mm}
\epsfysize=6.cm \epsfbox{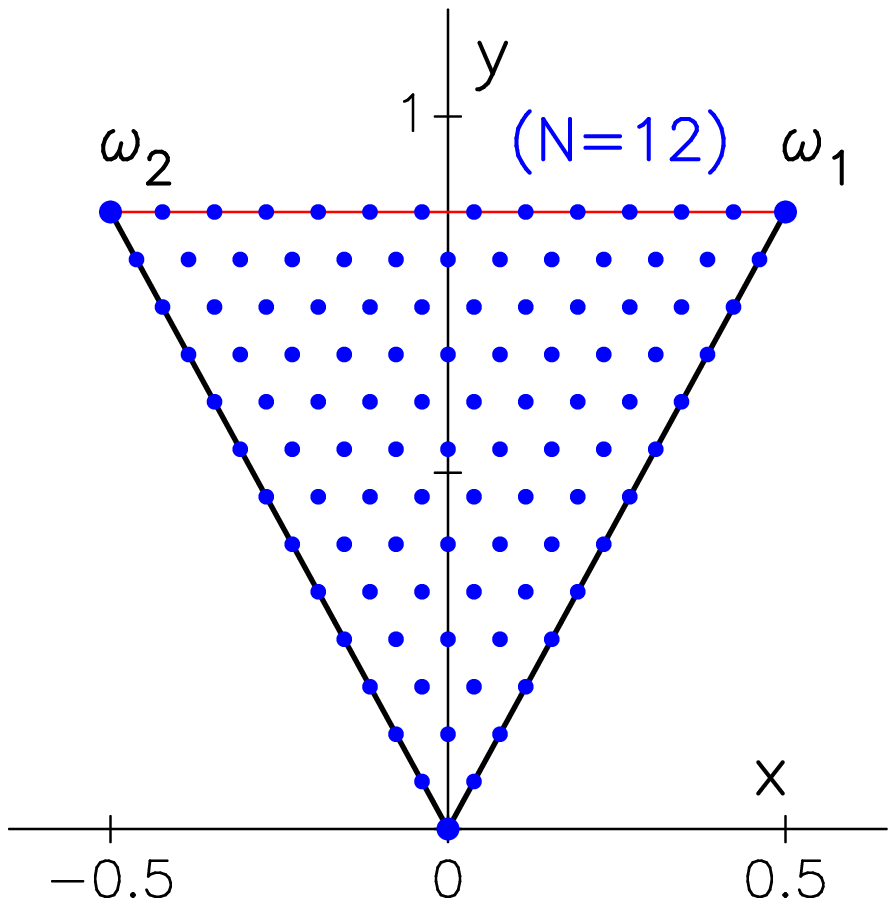}
}
\caption{
{\bf a }({\it left panel}):~~Representation of the maximal torus {\sf T} 
(hexagon) of SU(3) 
group in the plane of simple roots $\alpha_1$ and $\alpha_2$ of the group,
the fundamental domain {\sf F} and the fundamental weights $\omega_1$ 
and $\omega_2$. Also shown are the Weyl group orbits of two elements
in {\sf F}, which produce Weyl groups consisting of
 6 elements (shown by stars) or 3 elements (big dots);  
 {\bf b }({\it right panel}):~~ A set of  
elements of finite adjoint order $N=12$ (big dots)
in the fundamental domain (the triangle). 
}
\end{figure}

\begin{figure}
\centerline{
\epsfxsize=12.cm \epsfbox{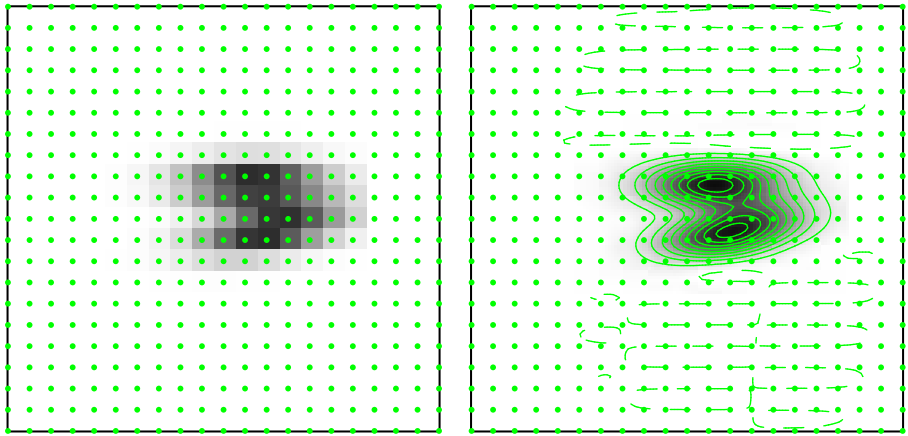}
}
\caption{{\bf a} ({\it left panel})~~ The discrete image  
produced from sampling of a continuous function $G(r)$  
composed of two Gaussian ellipsoids with large axes inclined at an angle
25$^\circ$ to each other, and with widths corresponding to dispersions $\sigma_\perp = 2/3N$ for a rectangular grid with $N=20$; {\bf b} ({\it right panel}):~~
The reconstructed CEDGT image. The contours correspond to intensity levels 
 separated by  
$10\%$ of the peak intensities in the ellipsoids, 
starting from the dashed contour 
at a small negative level of $-0.1\%$. 
}
\end{figure}

\clearpage

\begin{figure}
\centerline{
\epsfxsize=10.cm \epsfbox{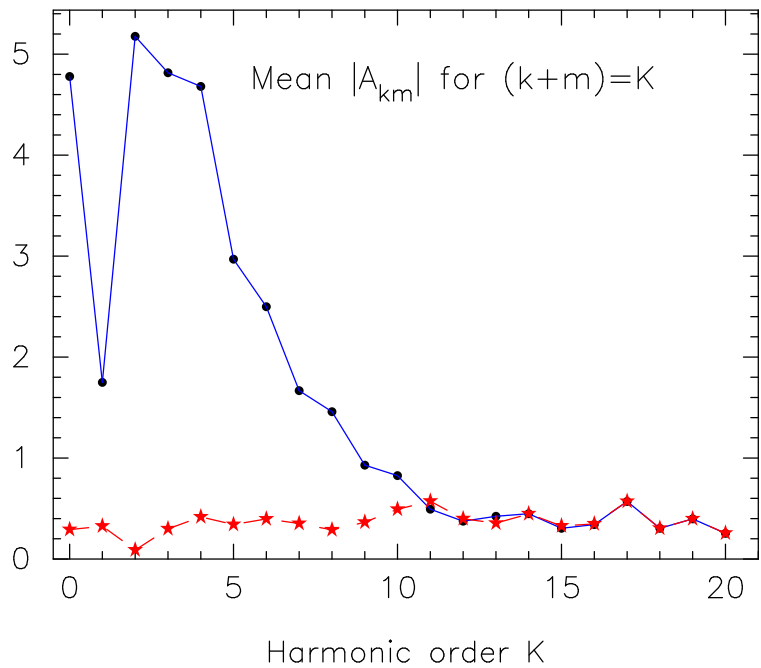}\hspace{5mm}
}
\caption{
The averaged absolute values $\mid A_{km} \mid$ of the DGT coefficients 
of the given harmonic order $K = k+m$ (shown by big dots)
calculated for the image presented in Fig.~4. The stars show 
the mean values $\mid A_{km}^\prime \mid$ of the transform of the
2 'hot pixels' (see the caption in Fig.~4) alone.}
\end{figure}

\begin{figure}
\centerline{\epsfig{file= 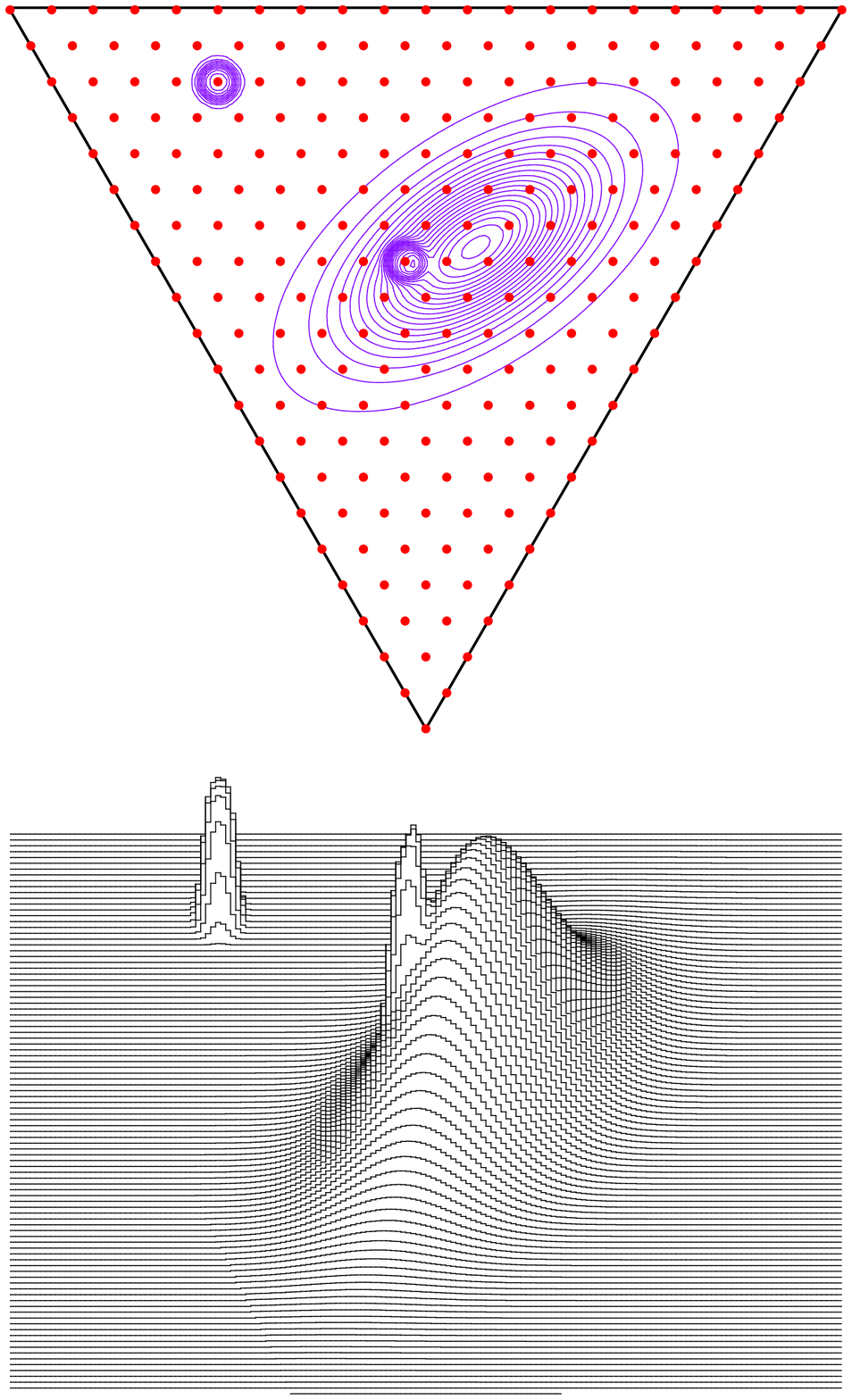, width=5.5cm}
      \epsfig{file= 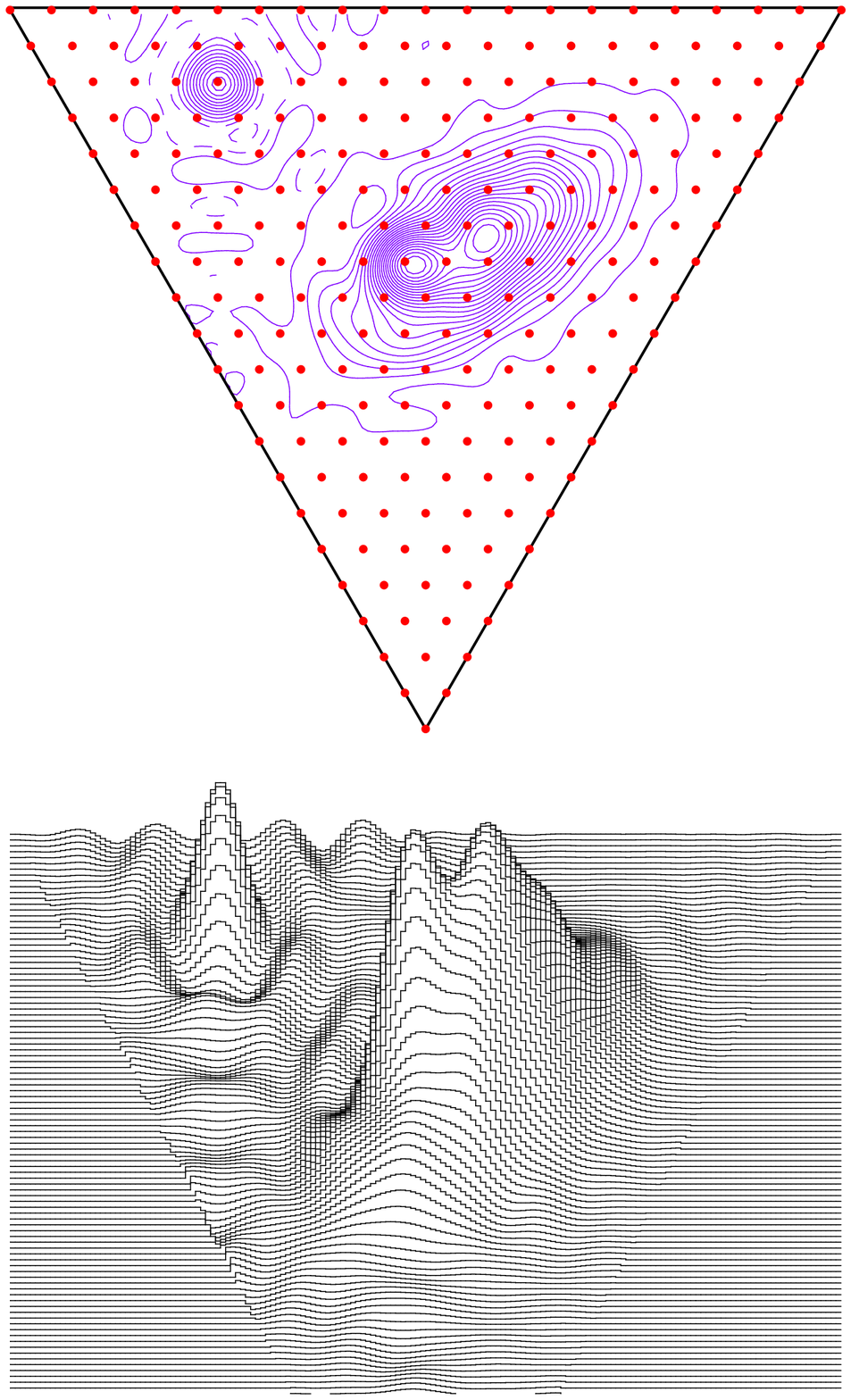, width=5.5cm}
       \epsfig{file= 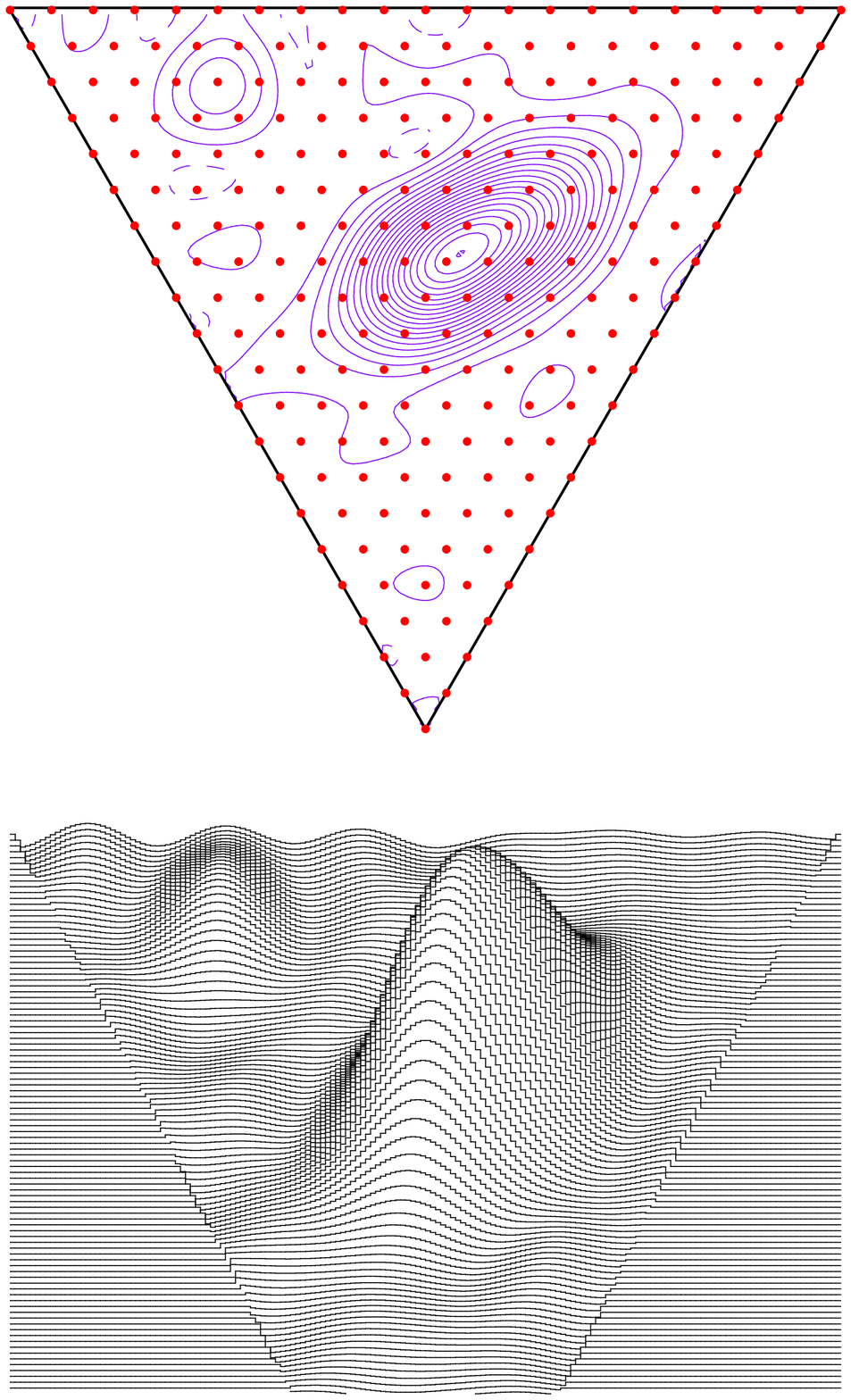, width=5.5cm}}
\caption{{\bf a}~({\it 2 panels on the left}):~~The contour plot
(upper panel) and the 3-dimensional view of an analog image
composed of 2-dimensional continuous Gaussian ellipsoid, onto which
2 narrow spikes which produce 2 `hot pixels' (after sampling of 
the analog signal on the grid) are superimposed. The heights of the 
spikes are equal to half of the height of the ellipsoid;~~{\bf b}~ 
({\it 2 panels in the middle}):~~A direct (i.e. {\it without filtering})
CEDGT interpolation of the 
discrete image produced by sampling of the analog image on the grid
with $N=20$;~~{\bf c}~({\it 2 panels on the right}):~~CEDGT view of 
the image after application of the filter with $C_f=0.5$ 
(see text).
}
\end{figure}

\clearpage

\begin{figure}
\centerline{
\epsfysize=5.5cm \epsfbox{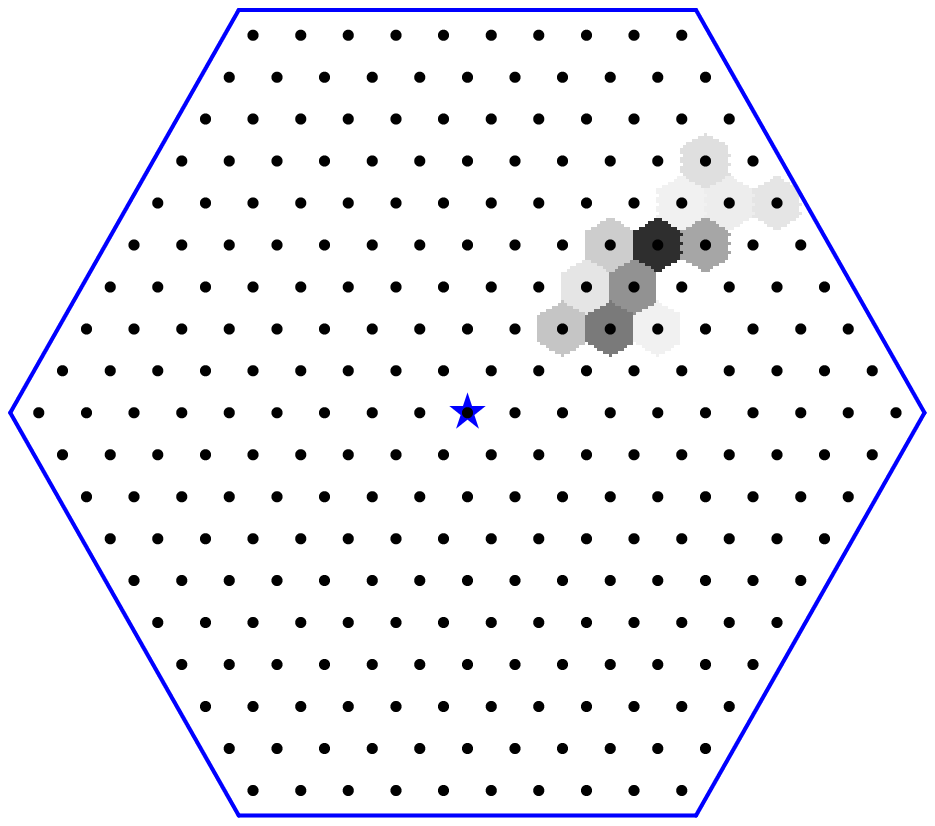}
\epsfysize=5.5cm \epsfbox{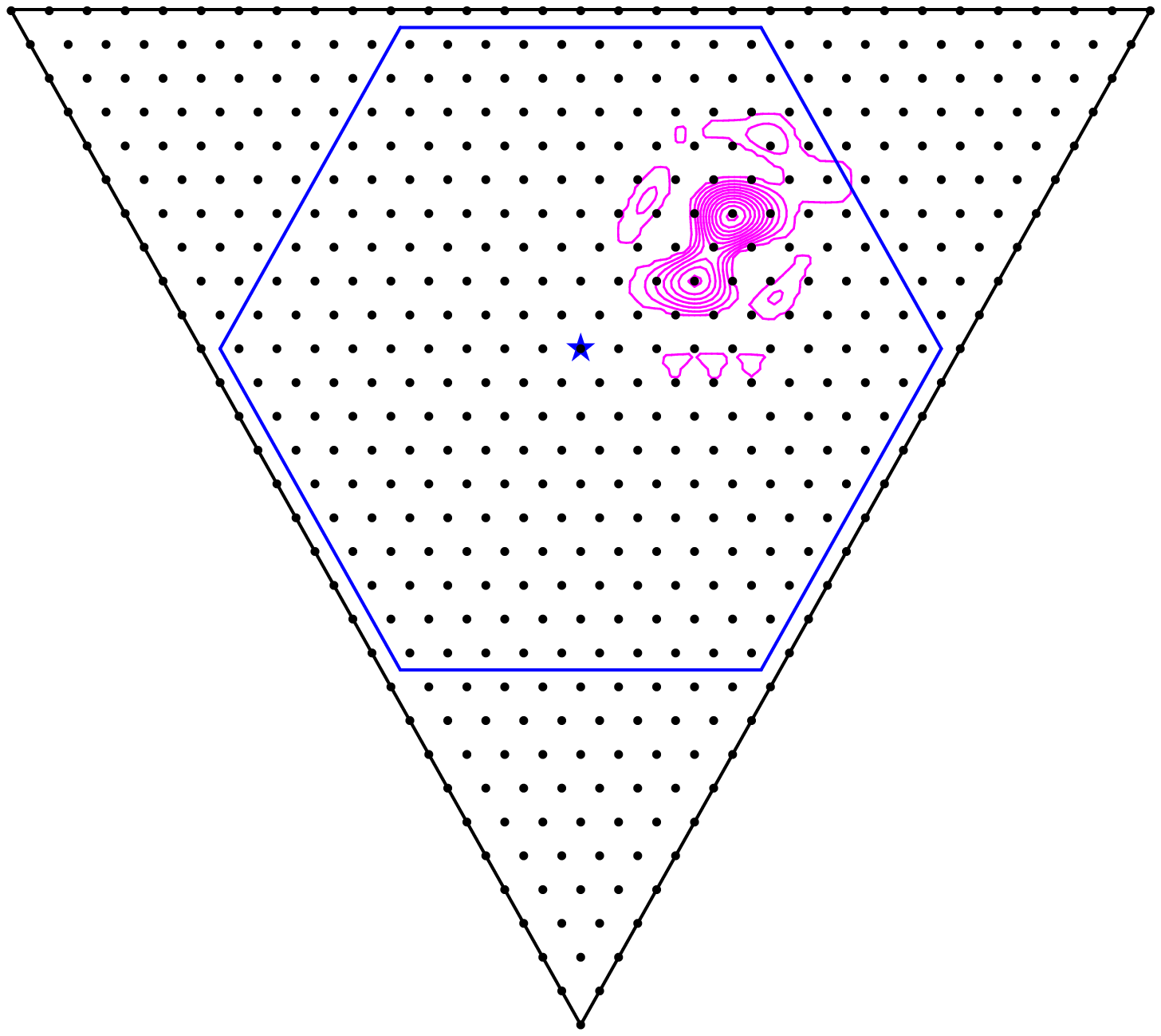}
\epsfysize=5.5cm \epsfbox{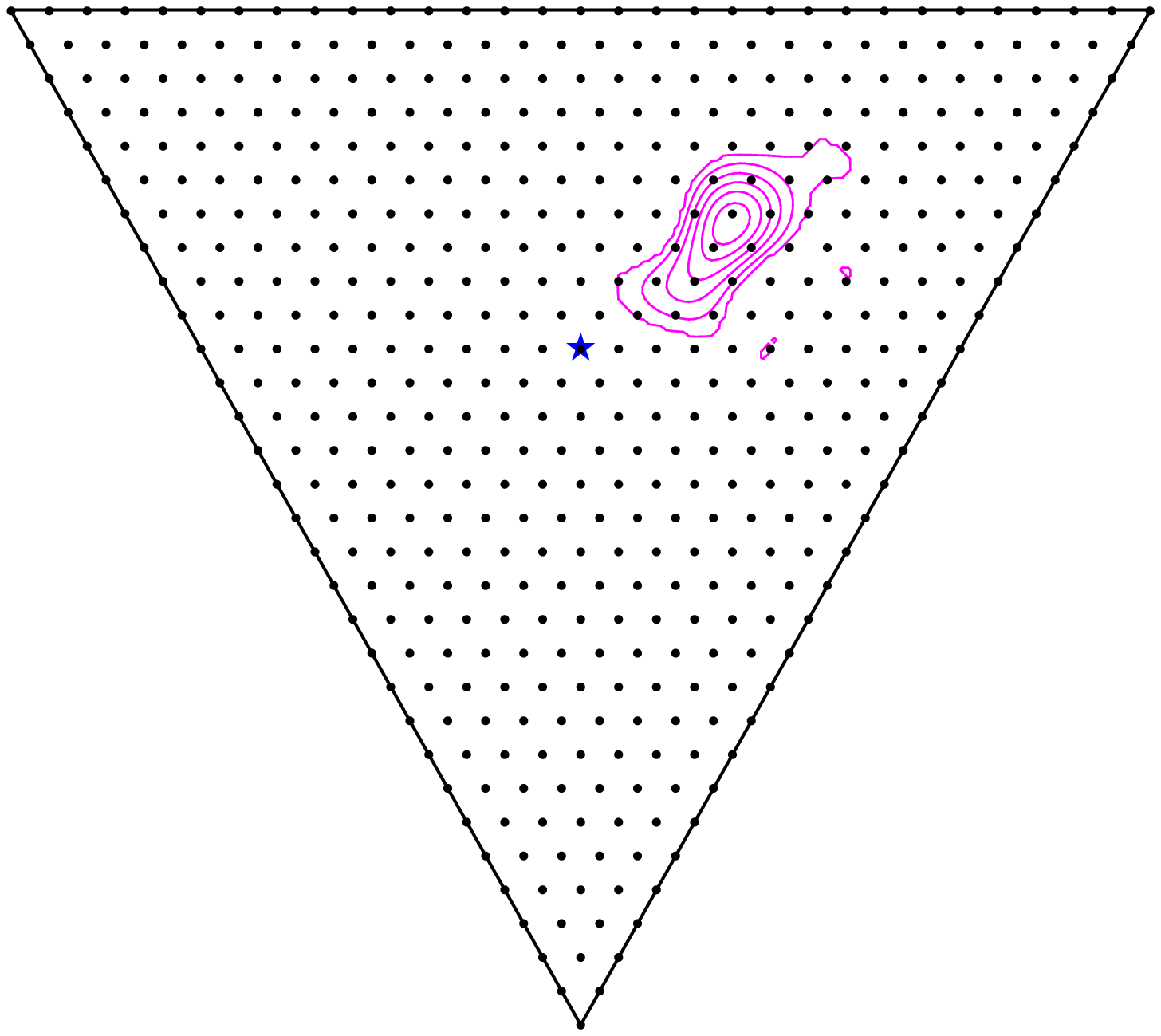}
}
\caption{{\bf a} ({\it left panel}):~~The discrete image of a gamma-ray
event with energy $E=1\,\rm TeV$ falling at the distance 130\,m from
the telescope;~~{\bf b}~~The contour plot of the same event in the CEDGT
representation without filtering, $C_f=0$. The lowest contour shown
corresponds to the level of $3\,\%$ from the maximum intensity of the 
image. The camera embedded into the fundamental triangle is also shown;
~~{\bf c}~~The image of the same EAS after application of the low-pass
filter with the parameter $C_{f} = 0.35$ (see text).
}
\end{figure}

\begin{figure}
\centerline{
\epsfysize=5.5cm \epsfbox{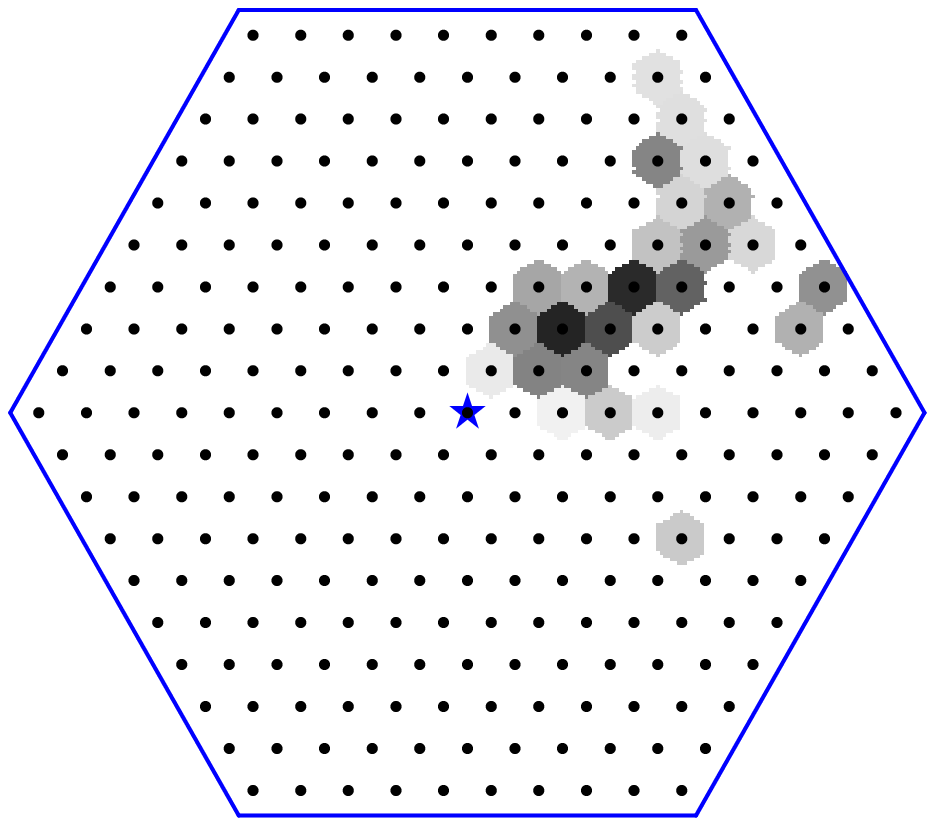}
\epsfysize=5.5cm \epsfbox{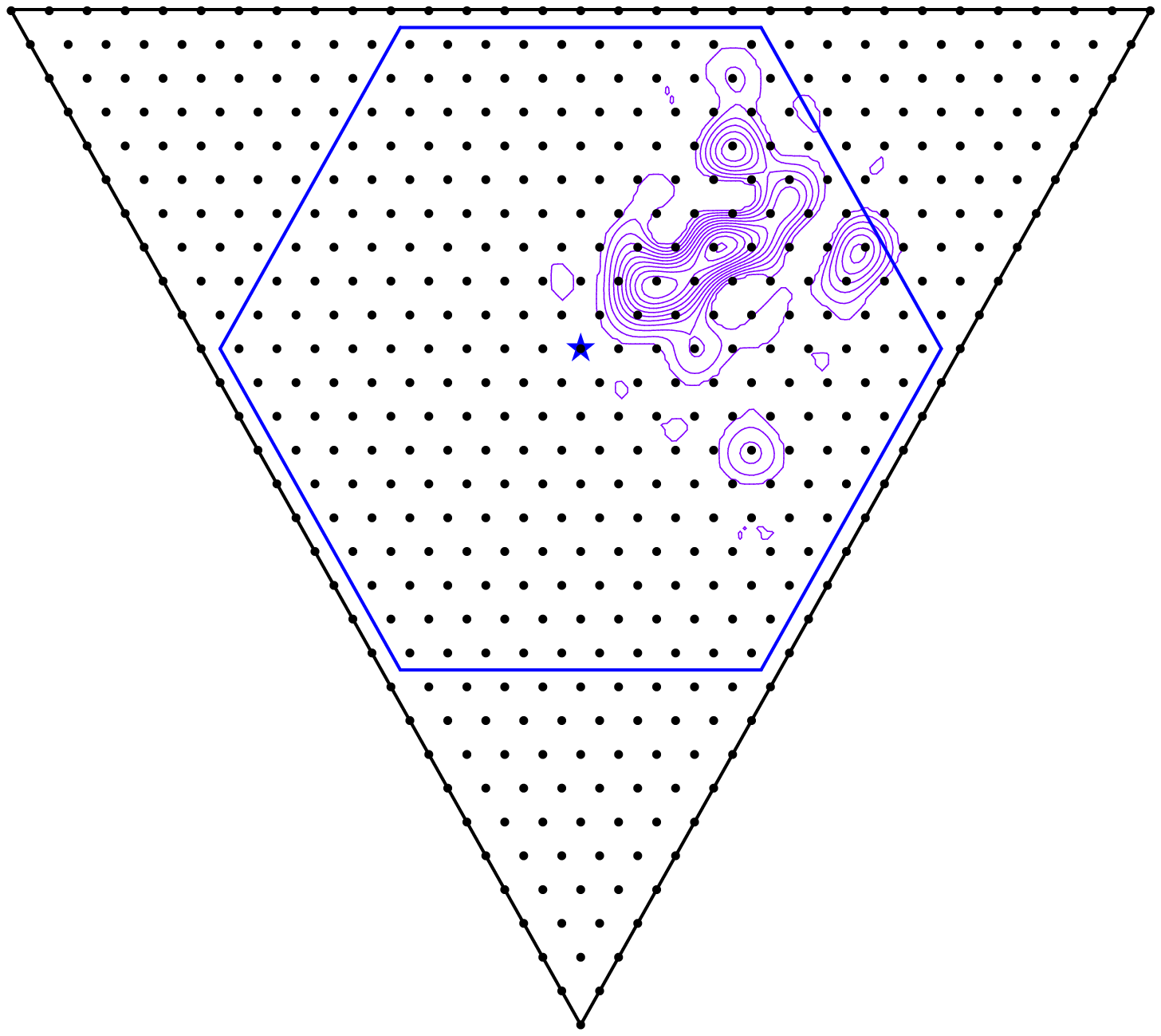}
\epsfysize=5.5cm \epsfbox{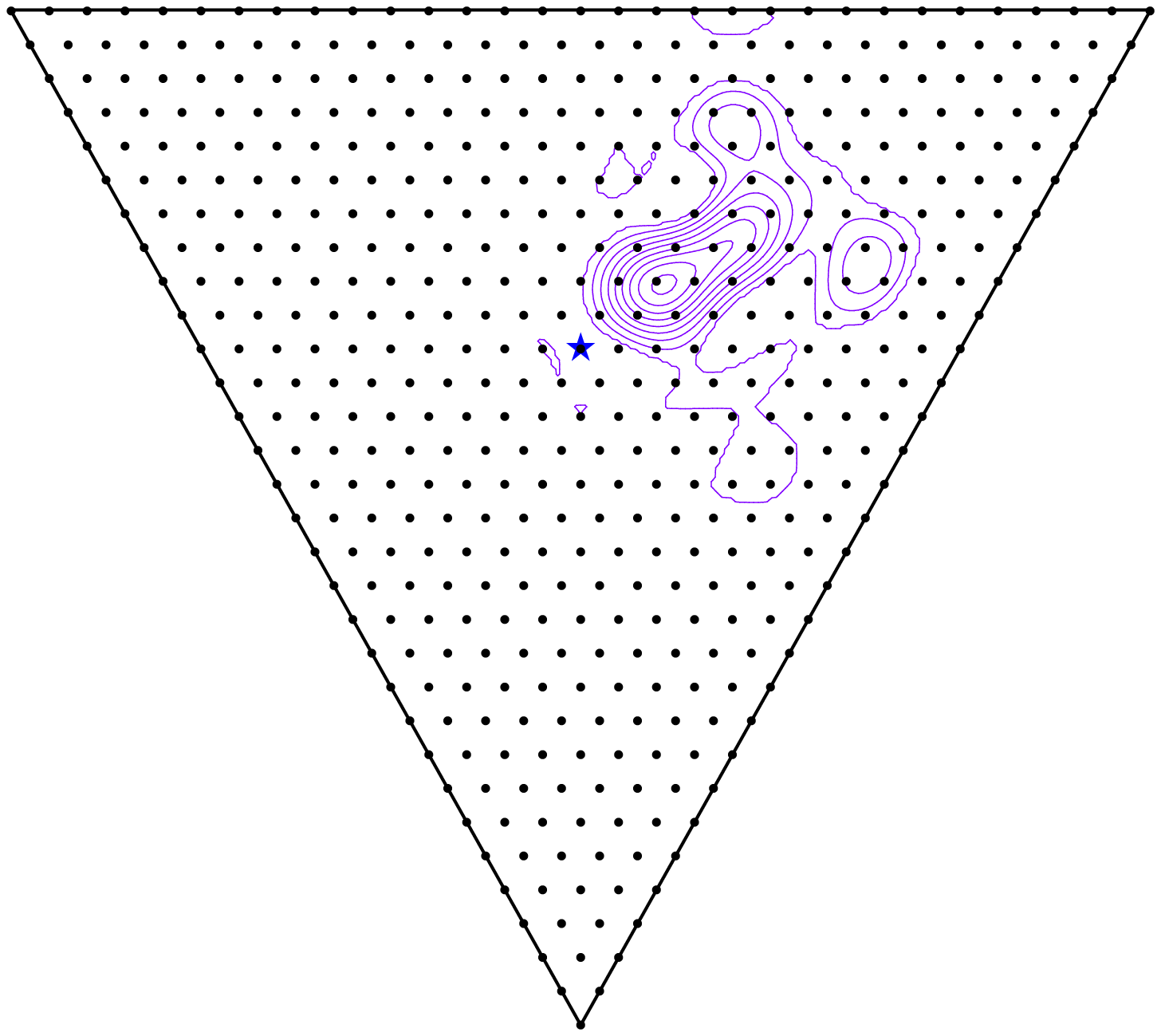}
}
\caption{
The same as in Figure 5, but for the proton EAS with energy  2 TeV
incident at a distance 100\,m from the telescope.
}
\end{figure}

\clearpage

\begin{figure}
\centerline{
\epsfig{file= 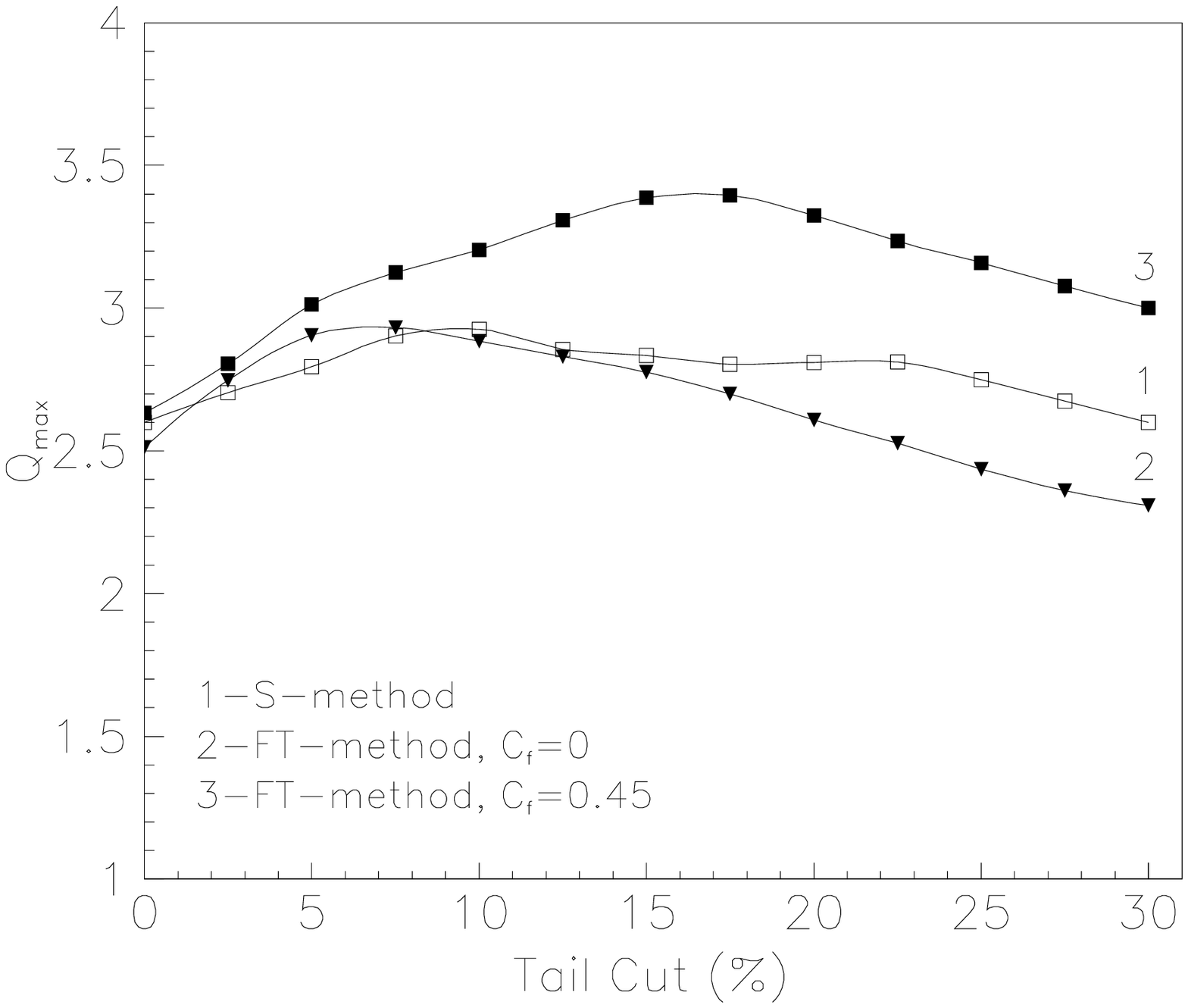, width=8.cm}
      \epsfig{file= 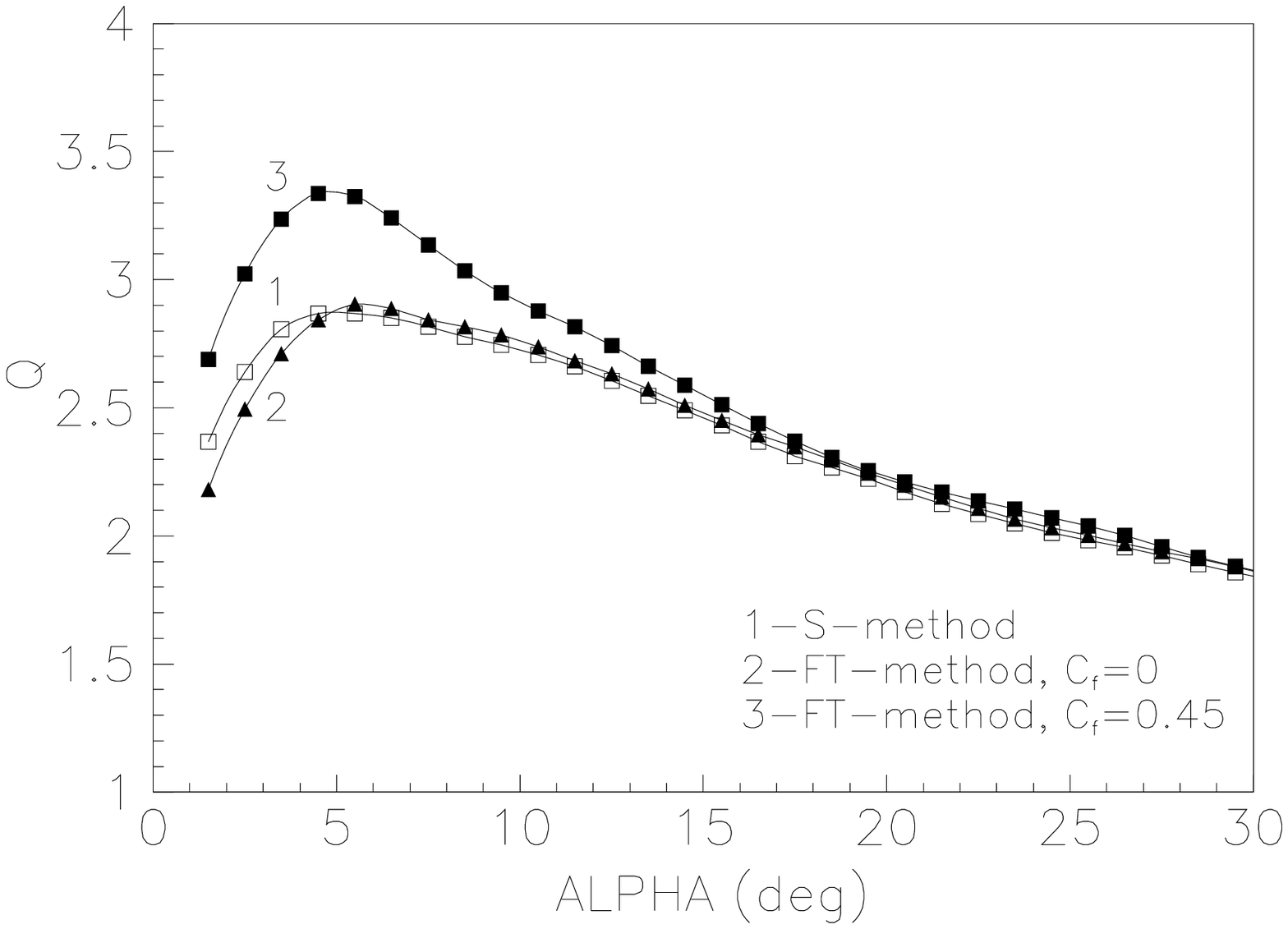, width=8.cm, height=6.8cm}
}
\caption{{\bf a } ({\it left panel}):~~The maximal $Q$-factors 
attained at different tail-cuts in the standard approach (curve 1, open 
squares), and by the Fourier transform method without filtering (curve 2,
triangles), and in case of low-pass filter with the parameter $C_f = 0.45$
(curve 3, full squares); 
~~{\bf b} ({\it right panel}):~~The depences of the $Q$-factors on $\alpha$
for the same 3 cases, but when the tail-cuts  
are fixed at the respective values corresponding to the 
absolute maxima of $Q_{max}$ of the curves on the left panel.
}
\end{figure}

\begin{figure}
\centerline{
\epsfxsize=8.cm \epsfbox{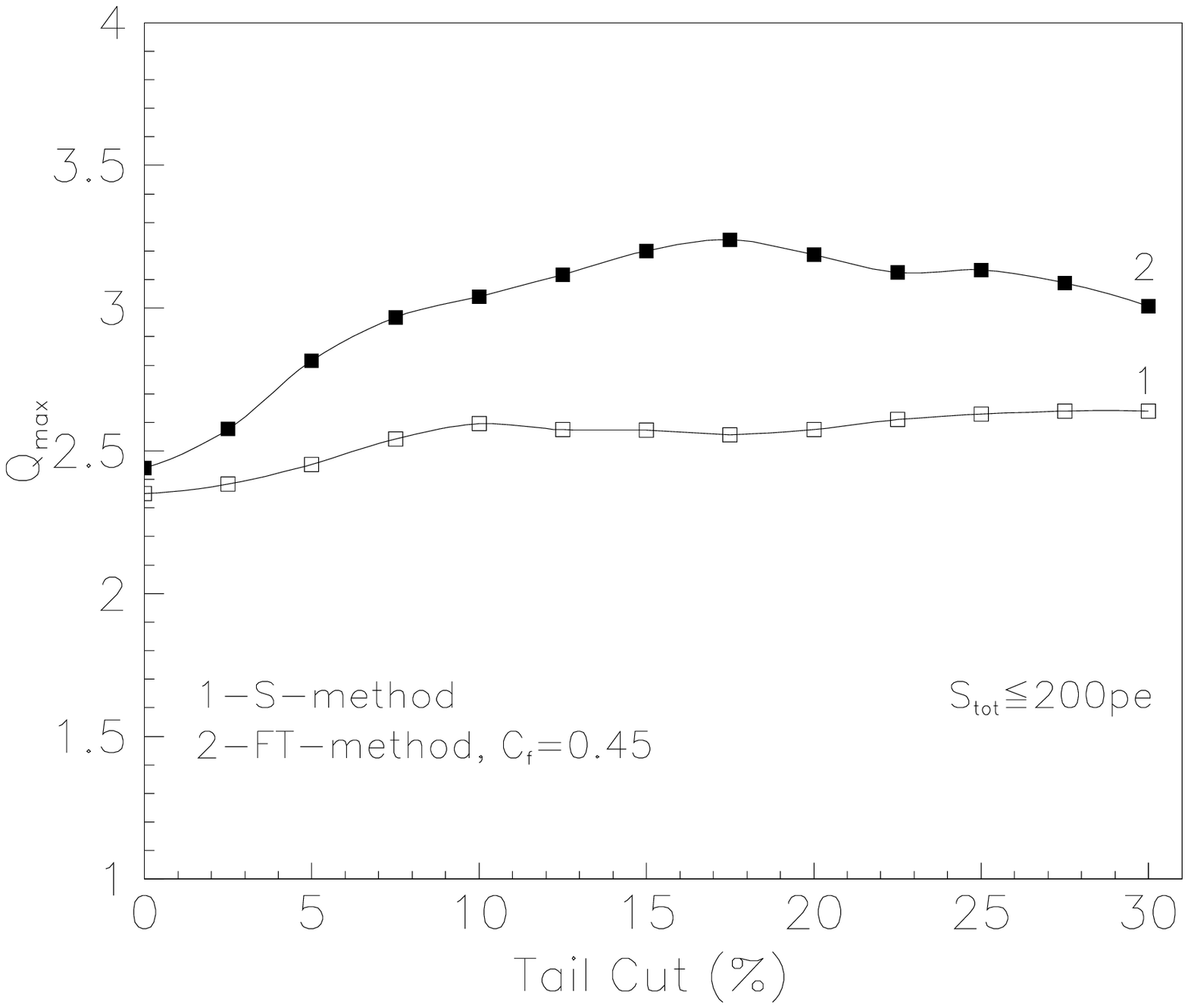} \hspace{1mm}
\epsfxsize=8.cm \epsfbox{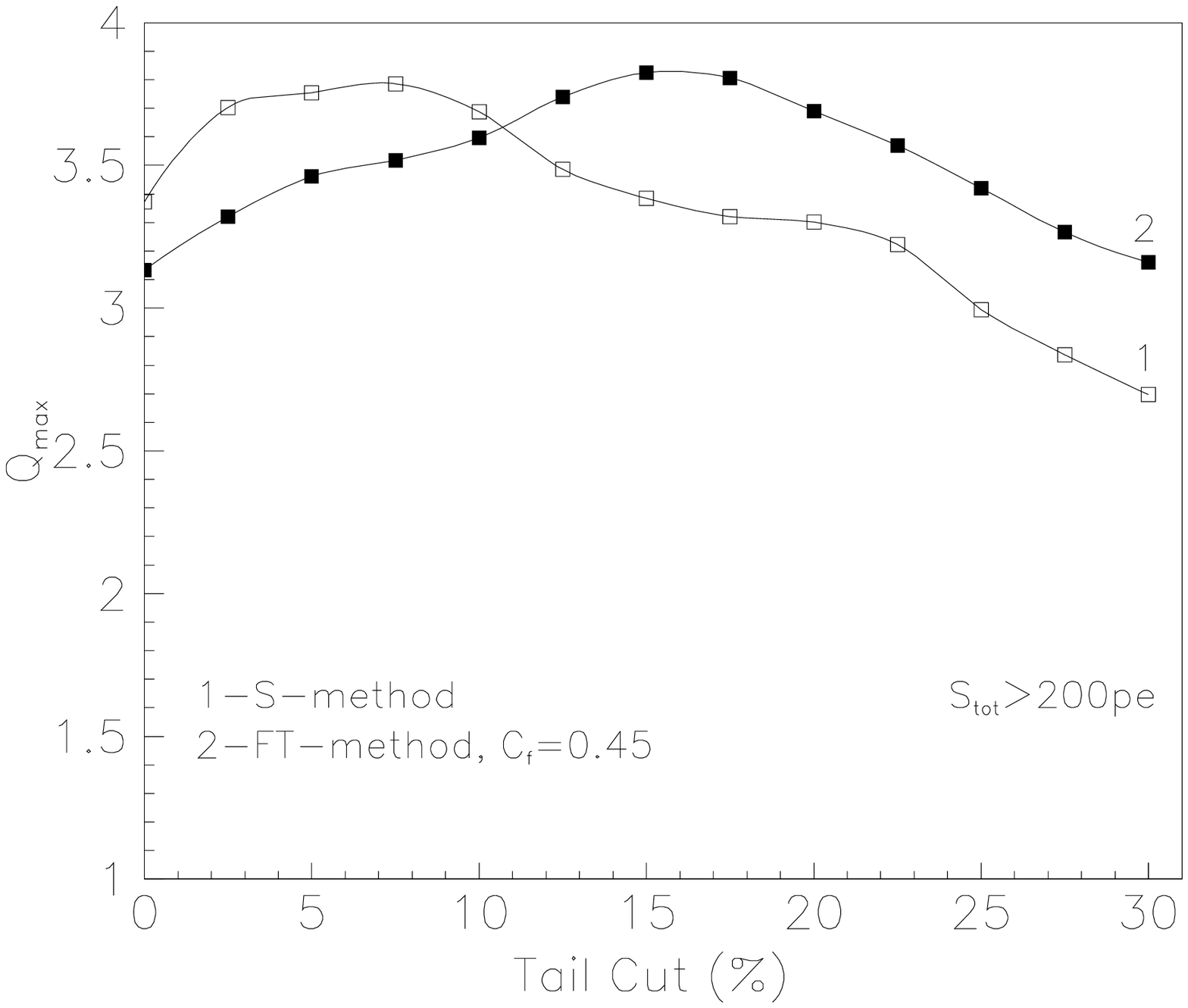}
}
\caption{The maximal $Q$-factors obtained by the S-method (open squares) and 
by the FT-method using the filter $C_f=0.45$, calculated for 
2 subsets of the original M-C image bank (used for Figure 7) that  
contain only photon-poor images with $S_{tot}\leq 200$ 
({\bf a} - {\it left panel}), or only  photon-rich images with 
$S_{tot} > 200$  ({\bf b} - {\it right panel}).
}
\end{figure}


\clearpage

\begin{figure}
\centerline{
\epsfxsize=8.cm \epsfbox{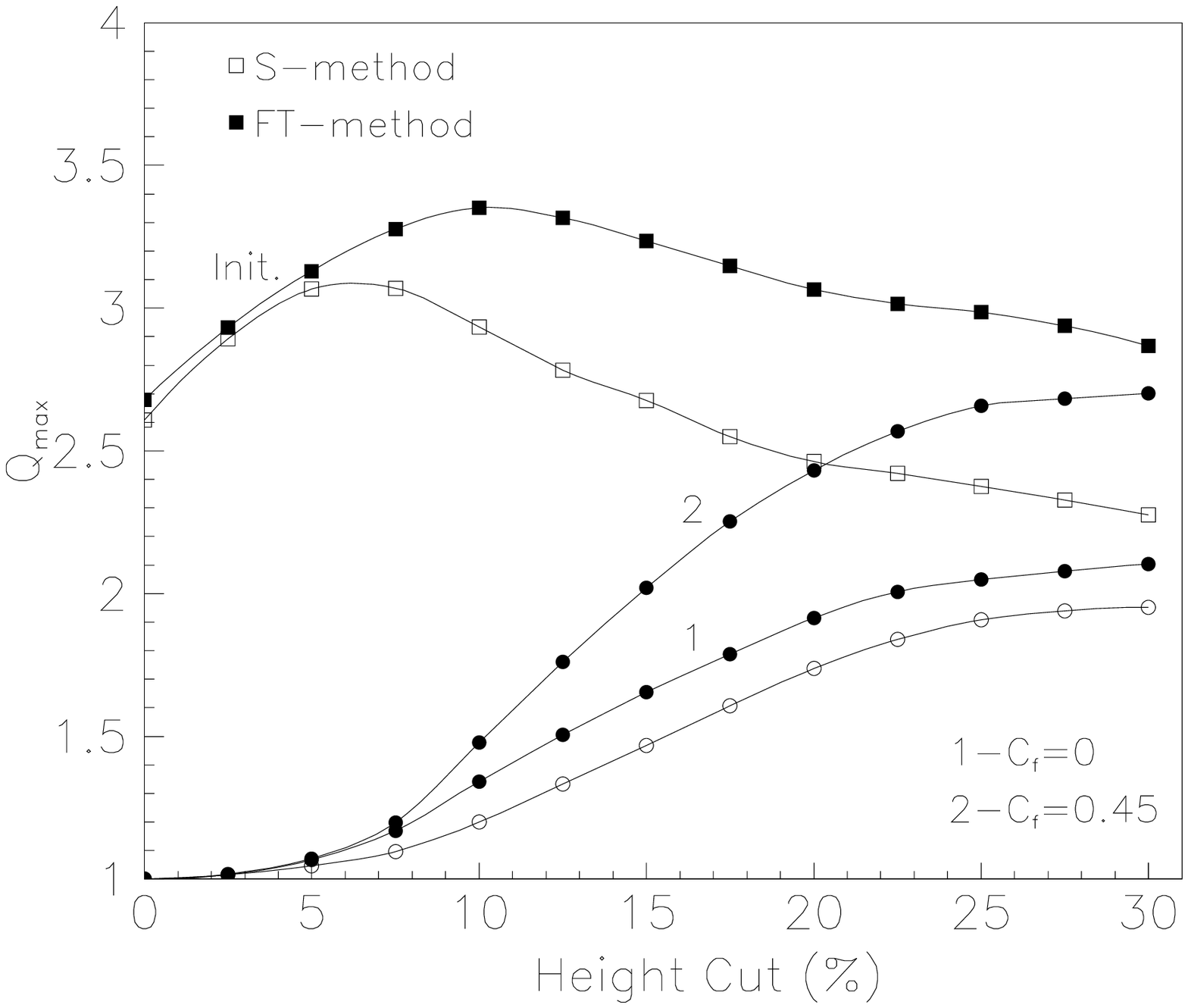} \hspace{1mm}
\epsfxsize=8.cm \epsfbox{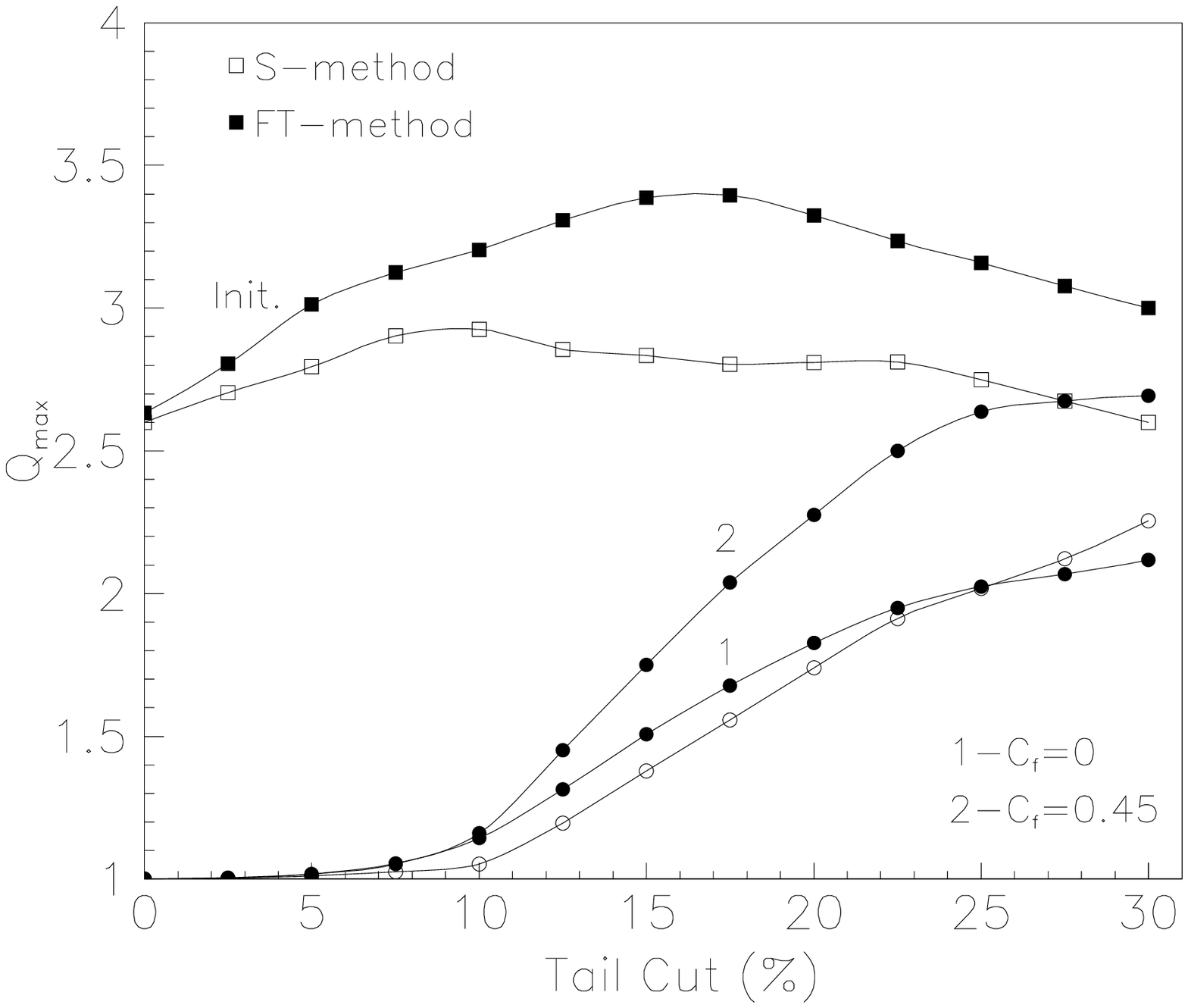}
}
\caption{
The recovery of the Q-factors with increasing levels of height cuts
({\bf a} - {\it left panel}) or tail cuts ({\bf b} - {\it right panel})
for the noisy data sets in case of the S-method (open circles),
and the  FT-method (full circles) without filtering (curve 1)
and using the filter with $C_f=0.45$ (curve 2). For comparison, the 
open and full squares show the Q-factors for the initial (pure) 
image set. For the noisy images pedestal substruction equal to $2\,pe$
has been used (see text).
}
\end{figure}

\begin{figure}
\centerline{
\epsfxsize=8.cm \epsfbox{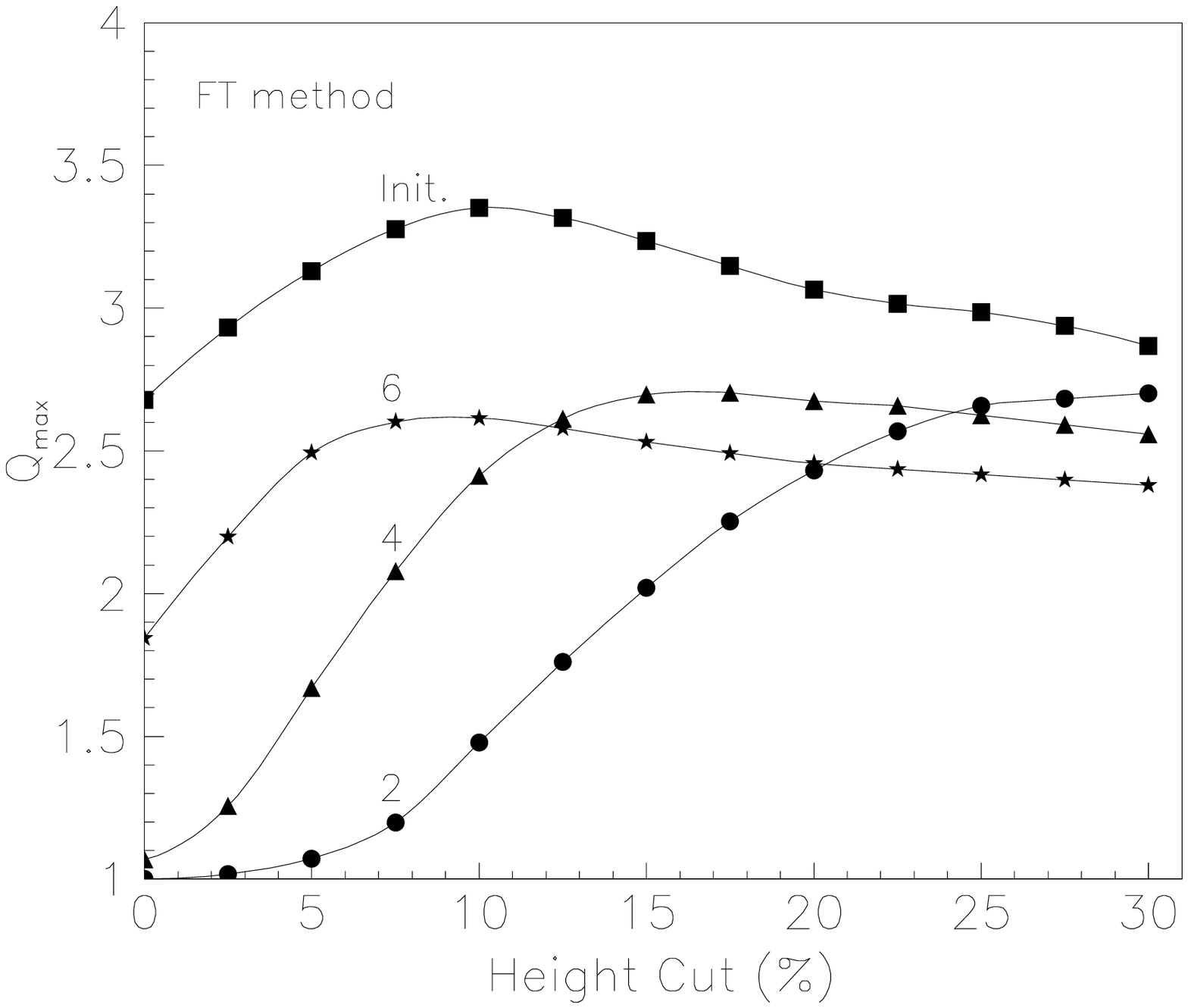} \hspace{1mm}
\epsfxsize=8.cm \epsfbox{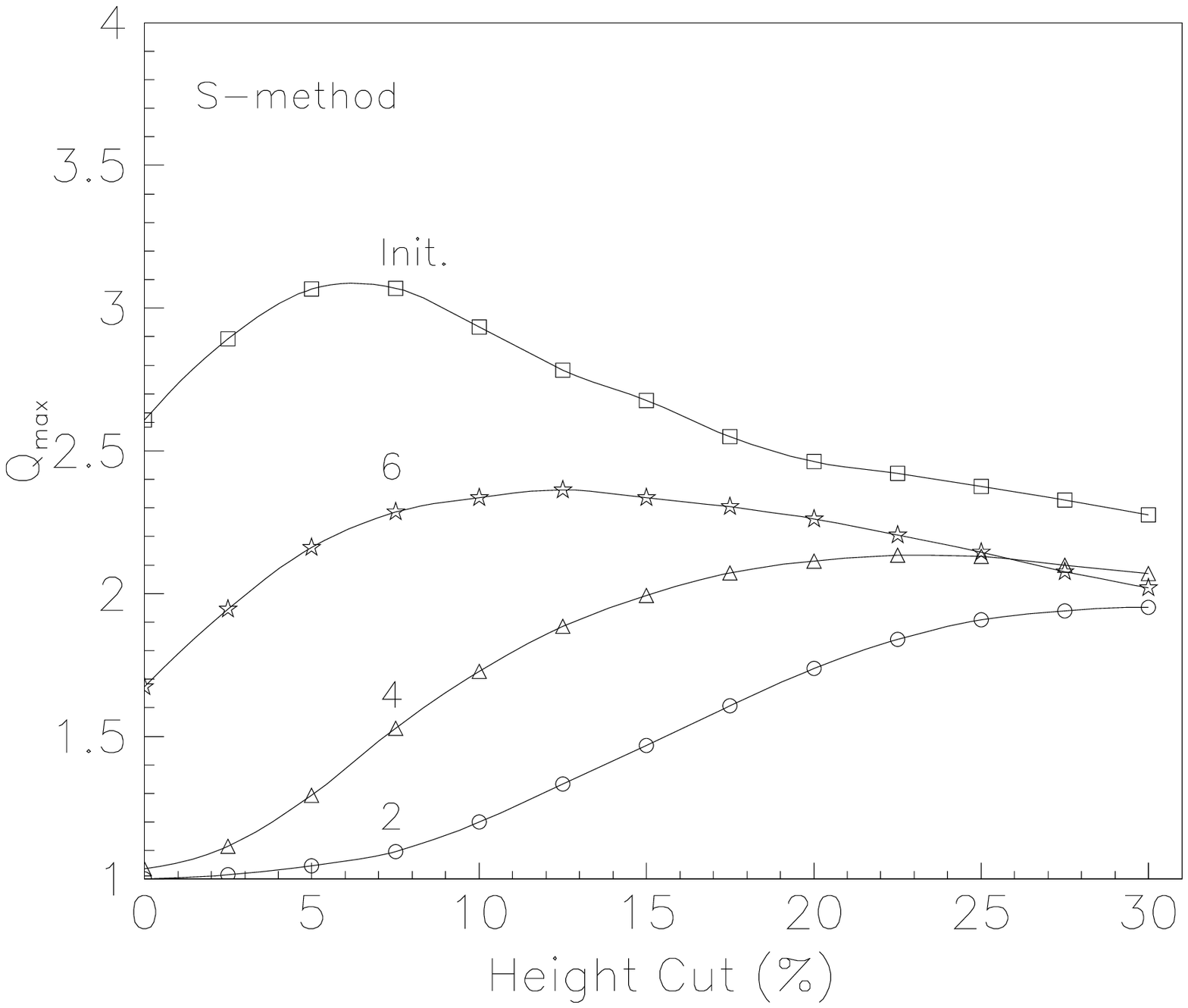}
}
\caption{
The extents of recovery of the signals from the noisy data sets 
with different levels of pedestal substruction (see text) in case 
of FT-method ({\bf a} - {\it left panel}) and S-method ({\bf b} - 
{\it right panel}). For comparison, for both methods also the initial 
Q-factors corresponding to the set of pure images are presented.
}
\end{figure}


\begin{thebibliography}{999}
\bibitem{crab89} T.\,C.\ Weekes,  et al., {\em ApJ} {\bf 342} (1989), 379.
\bibitem{AhAk97}
 F. A. Aharonian, and C. W. Akerlof, {\em Ann. Rev. Nucl. Part. Sci.} 
    {\bf 47} (1997), p. 273. 
\bibitem{TW03}
T. C. Weekes, {\em Very high energy gamma-ray astronomy}, The Institute of
                     Physics Publishing, (Bristol, UK, 2003)
\bibitem{tw78}
K. E. Turver, and T. C. Weekes, {\em Nuovo Cim.} {\bf 45B} (1978), 78.  
\bibitem{hillas}
 A. M. Hillas, in: Proc. 19th ICRC (La Jolla), Vol. 3 (1985), p. 445.
\bibitem{mohanty}
  G. Mohanty, et al., {\em Astropart. Phys.} {\bf 9} (1998), 14.
\bibitem{lessard}
 R. W. Lessard, et al. {\em Astropart. Phys.} {\bf 17} (2002), 417.
\bibitem{bhb}
I. H. Bond, A. M. Hillas, and S. M. Bradbury, 
   {\em Astropart. Phys.} {\bf 20} (2003), 311.
\bibitem{oppenh}
A. V. Oppenheim, and R. W. Schafer, {\it Digital signal processing},   
 Englwood Cliffs, Prentice-Hall, (1975)
\bibitem{lj90}
J. S. Lim,  {\em Two-dimensional signal and image processing}, 
Englewood Cliffs, N.J., Prentice Hall (1990).
\bibitem{MP1}
R. V. Moody and J. Patera, {\em SIAM J. on Algebraic and Discrete Methods}
{\bf 5} (1984), 359. 
\bibitem{MP2}
R. V. Moody, and J. Patera, {\em Mathematics of Computation} {\bf 48} (1987),
799.
\bibitem{P03}
J. Patera, in CRM Proc. ``Group Theory and Numerical Methods'' 
(Montreal, 26-31 May, 2003), eds. D. Gomez-Ullate, et al., CRM Proc. series,
to be published.
\bibitem{CEDCT}
A. Atoyan, and J. Patera,  {\em J. Math. Phys.} {\bf 45} (2004), 2468.
\bibitem{DCT}
K. R. Rao, and P. Yip, {\it Discrete cosine transform Algorithms, 
Advantages, Applications}, Academic Press (1990) 
\bibitem{hess}
W. Hofmann, in: M. Simon, E. Lorenz, M. Pohl (Eds.), 
Proc. of the 27th Int. Cosmic Ray Conf., Hamburg, 2001, p. 2785.
\bibitem{MAGIC}
E. Lorenz, The MAGIC Collaboration. in ``GeV-TeV Gamma Ray Astrophysics 
Workshop: towards a major atmospheric Cherenkov detector VI'', 
(Snowbird, Utah, 13-16 August 1999), eds. B. L. Dingus, M. H. Salamon, and 
D. B. Kieda. Melville, N.Y.: AIP, 2000. AIP Conf. Proc., vol. 515., p.510
\bibitem{veritas}
T.C. Weekes, et al., {\em Astropart. Phys.} {\bf 17} (2002),  221.
\bibitem{cangaroo3}
S. Kabuki, et al., {\em NIM} {\bf A500} (2003), 318.
\bibitem{cangaroo2}
A. Kawachi, et al., {\em Astropart. Phys.} {\bf 14} (2001),  261.
\bibitem{ALPHA}
D. J. Fegan, {\em J. Phys.} {\bf G 23} (1997), 1013.
\bibitem{HEGRA}
    G. P\"uhlhofer, et al., {\em Astropart. Phys.} {\bf 20} (2003), 267.
\bibitem{MOCCA}
A. Hillas, {\em Nucl.Phys. B. (Proc. Suppl.)} {\bf 52B} (1997), 29.
\bibitem{rtracing}
 A. Akhperjanian, et al.,  {\em Exp. Astron.} {\bf 8} (1998), 135.


\end{thebibliography}
\end{document}